%% file: main.tex
\documentclass[sigconf,numbers]{acmart}
\setcopyright{none}
\renewcommand\footnotetextcopyrightpermission[1]{}

\input{_includes}

\AtBeginDocument{
\providecommand\BibTeX{{
    \normalfont B\kern-0.5em{\scshape i\kern-0.25em b}\kern-0.8em\TeX}}}

\begin{document}

\input{macro}

\title[\projectName{} Generating Simulated Video Comments through Multimodal AI and Personas]{\projectName{}: Generating Simulated Video Comments through Multimodal AI and User Personas
}

\author{Yu-Kai Hung}
\email{b09902040@csie.ntu.edu.tw}
\orcid{orcid}
\affiliation{
  \institution{National Taiwan University}
  \streetaddress{No. 1, Sec. 4, Roosevelt Rd.}
}

\author{Yun-Chien Huang}
\email{b09902067@csie.ntu.edu.tw}
\orcid{orcid}
\affiliation{
  \institution{National Taiwan University}
  \streetaddress{No. 1, Sec. 4, Roosevelt Rd.}
}

\author{Ting-Yu Su}
\email{tingyusu1786@gmail.com}
\orcid{orcid}
\affiliation{
  \institution{National Taiwan University}
  \streetaddress{No. 1, Sec. 4, Roosevelt Rd.}
}

\author{Yen-Ting Lin}
\email{ytl@ieee.org}
\orcid{orcid}
\affiliation{
  \institution{National Taiwan University}
  \streetaddress{No. 1, Sec. 4, Roosevelt Rd.}
}

\author{Lung-Pan Cheng}
\email{lung-pan.cheng@csie.ntu.edu.tw}
\orcid{orcid}
\affiliation{
  \institution{National Taiwan University}
  \streetaddress{No. 1, Sec. 4, Roosevelt Rd.}
}

\author{Bryan Wang}
\authornote{Equal advisory contribution. \\ Correspondence to: Shao-Hua Sun \texttt{<shaohuas@ntu.edu.tw>}}
\email{bryanw@dgp.toronto.edu}
\orcid{orcid}
\affiliation{
  \institution{University of Toronto}
  \streetaddress{}
}

\author{Shao-Hua Sun}
\authornotemark[1]
\email{shaohuas@ntu.edu.tw}
\orcid{orcid}
\affiliation{
  \institution{National Taiwan University}
  \streetaddress{No. 1, Sec. 4, Roosevelt Rd.}
}

\renewcommand{\shortauthors}{Hung et al.}

\begin{abstract}
Audience feedback is crucial for refining video content, yet it typically comes after publication, limiting creators' ability to make timely adjustments. To bridge this gap, we introduce \projectName{}, a generative AI system designed to simulate audience feedback in the form of video comments before a video's release. \projectName{} features a computational pipeline that integrates multimodal data from the video—such as visuals, audio, and metadata—with user personas derived from a broad and diverse corpus of audience demographics, generating varied and contextually relevant feedback. Furthermore, the system’s UI allows creators to explore and customize the simulated comments. Through a comprehensive evaluation—comprising quantitative analysis, crowd-sourced assessments, and qualitative user studies—we show that \projectName{}'s generated comments are not only relevant, believable, and diverse but often more detailed and informative than actual audience comments, highlighting its potential to help creators refine their content before release.
\end{abstract}

\begin{teaserfigure}
\includegraphics[width=\textwidth]{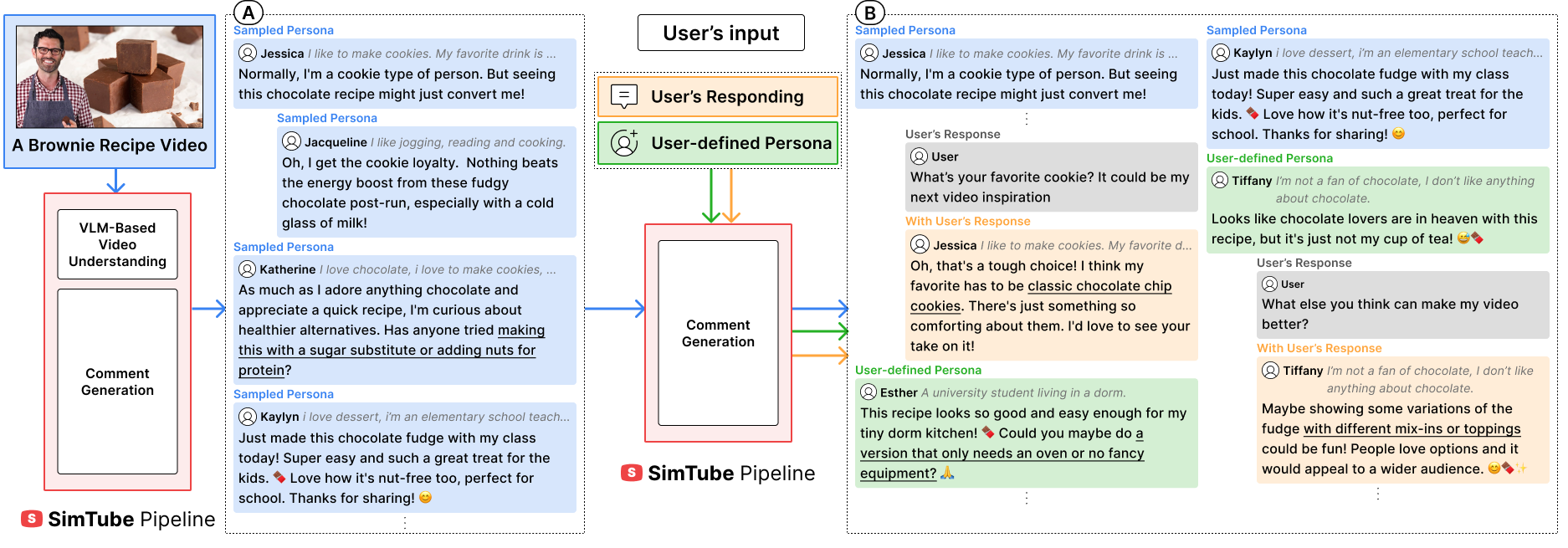}
    \caption{\emph{\projectName{}} (A) generates synthetic content-based video comments and (B) supports user-steerable comment generation from users' responding or defined persona, providing preliminary inspiration and insight for iterating video.}
  \Description{SimTube generates synthetic content-based video comments and supports user-steerable comment generation from users' responding or defined persona, providing preliminary inspiration and insight for iterating video.}
  \label{fig:figure1}
\end{teaserfigure}

\maketitle
\pagestyle{plain}

\input{sections/1_introduction}
\input{sections/2_related_works}
\input{sections/3_design_goals}
\input{sections/4_ui_Implementation}
\input{sections/5_backend_implementation}
\input{sections/6_evaluation}
\input{sections/7_user_study}
\input{sections/8_limitations}

\input{sections/9_conclusion}

\clearpage
\appendix
\input{sections/10_Appendix}

\begin{acks}
Shao-Hua Sun was supported by the Yushan Fellow Program by the Ministry of Education,
Taiwan.
\end{acks}

\bibliographystyle{ACM-Reference-Format}
\bibliography{reference}

\end{document}

%% file: _includes.tex
\usepackage{bm}
\usepackage{mathtools}
\usepackage{amsmath}
\usepackage{multirow}
\usepackage{xcolor}
\usepackage{float}

\usepackage{tabularx}  
\usepackage{booktabs}

\usepackage{multirow}
\usepackage{makecell}
\usepackage{stfloats}
\usepackage{siunitx}
\usepackage{comment}

\usepackage{blindtext}
\usepackage{soul}

\usepackage{graphicx}
\usepackage{color}
\usepackage{textcomp}
\usepackage{cuted}
\usepackage{enumerate}
\usepackage{enumitem}
\usepackage[all]{hypcap}    
\usepackage{subfigure}
\usepackage[utf8]{inputenc}
\usepackage{booktabs}
\usepackage{listings}

\usepackage[capitalize,noabbrev]{cleveref}
\usepackage{titlesec}
\titleclass{\subsubsection}{straight}
\titleformat{\subsubsection}%
  [hang]
  {\normalfont\large\itshape}
  {\thesubsubsection}
  {1em}
  {}
  []

%% file: macro.tex
\newcommand{\projectName}{SimTube}
\newcommand{\todo}[1]{\textcolor{red}{\textbf{<TODO: #1>}}} 
\newcommand{\draft}{\textcolor{blue}{<This part is drafted>}} 

\newcommand{\sun}[1]{{\color{blue}{\small\bf\sf [Sun: #1]}}}
\newcommand{\su}[1]{{\color{orange}{\small\bf\sf [Su: #1]}}}
\newcommand{\yk}[1]{{\color{orange}{\small\bf\sf [YuKai: #1]}}}

\newcommand{\blue}[1]{{\color{blue}#1}}
\newcommand{\red}[1]{{\color{red}#1}}
\newcommand{\task}[1]{{\color{red}{\small\bf\sf [TASK: #1]}}}
\newcommand{\Skip}[1]{}

\newcommand{\etal}{\textit{et al}.}
\newcommand{\ie}{\textit{i}.\textit{e}.,\ }
\newcommand{\eg}{\textit{e}.\textit{g}.,\ }

\newcommand{\myfig}[1]{Figure~\ref{#1}}
\newcommand{\mytable}[1]{Table~\ref{#1}}
\newcommand{\myeq}[1]{Eq.~\ref{#1}}
\newcommand{\mysecref}[1]{Section~\ref{#1}}
\newcommand{\myalgo}[1]{Algorithm~\ref{#1}}

\newcommand{\tieconcat}{\mathbin{\mathpalette\dotieconcat\relax}}
\newcommand{\dotieconcat}[2]{
  \text{\raisebox{.8ex}{$\smallfrown$}}%
}
\newcommand{\fix}{\marginpar{FIX}}
\newcommand{\new}{\marginpar{NEW}}

\newcommand{\myparagraph}[1]{\textbf{#1.}}
\newcommand{\vspacesection}[1]{\vspace{-0.00cm}
\section{#1}
\vspace{-0.0cm}}
\newcommand{\vspacesubsection}[1]{\vspace{-0.0cm}
\subsection{#1}
\vspace{-0.0cm}}
\newcommand{\vspacesubsubsection}[1]{\vspace{-0.0cm}
\subsubsection{\textbf{#1}}
\vspace{-0.0cm}}

\newcommand\blfootnote[1]{%
  \begingroup
  \renewcommand\thefootnote{}\footnote{#1}%
  \addtocounter{footnote}{-1}%
  \endgroup
}

\newcommand\SmallCaption[1]{%
  \captionsetup{font=scriptsize}%
  \caption{#1}}

%% file: sections/1_introduction.tex
\section{Introduction}
\label{sec:introduction}

Audience feedback is crucial for video content creators to shape and refine content. 
Various modern platforms~\cite{Chelaru2014socialfeedback,videocrit2016pavel,Ramos2003annotateVideo} allow creators to share videos and gather crowd-sourced input~\cite{veed2024,thematic2024,2023madeonyoutube,videocrit2016pavel}. 
For instance, YouTubers and TikTok creators frequently adjust their content based on audience reactions and comments from previous episodes~\cite{Holmbom2015YoutuberStudy, DMI2024growYT}.
However, such audience feedback is often delayed, typically provided only \textbf{after} the content has been published, limiting creators' ability to make timely improvements and dynamically adjust the content. 
That said, creators must wait until their content is posted to receive input, which can only influence future episodes rather than the current one. 
For novice creators, this challenge is even more pronounced due to the low visibility of their content, making it difficult to gather substantial, actionable feedback.

To address these limitations, we aim to develop methods to enable video content creators to obtain diverse and meaningful audience feedback \textbf{before} publishing their videos. 
Specifically, we focus on simulating \textit{audience comments}, a prevalent form of feedback on video-sharing platforms. 
Video comments allow viewers to directly share their thoughts with creators, often sparking broader discussions as other users engage by liking or replying~\cite{thelwall2012commenting}. 
These engagements offer valuable perspectives and foster a sense of community~\cite{thelwall2012commenting,DUBOVI2020103939}, making video comments a rich source of feedback for creators.

To this end, we present \projectName{}, a full-stack AI system capable of generating diverse, relevant, and believable audience comments based on video content. 
We proposed a computational pipeline that leverages generative AI models, including vision language models (VLMs) for understanding visuals, speech recognition for transcribing audio, and large language models (LLMs) for generating natural language feedback. 
This pipeline first integrates the multimodal data in videos—including visuals, audio, and metadata—to produce a video summary and keywords.  
Subsequently, the video summary and keywords are combined with various persona descriptions—representing different audience demographics and backgrounds—to simulate video comments from diverse perspectives. In addition to this pipeline, we designed a user interface that allows creators to upload videos, obtain simulated feedback, and interactively explore the results. 
Users can also customize personas to tailor the feedback to specific viewpoints or audience backgrounds as they see fit.

To understand the effectiveness of \projectName{} and the quality of the comments it generates, we conducted a comprehensive set of assessments, including quantitative analysis, crowd-sourced ratings, and qualitative studies. Our results indicate that \projectName{} produces relevant, believable, and helpful comments for creators across various video genres. Notably, in many instances, AI-generated comments were rated as more informative and beneficial to creators than those left by actual users. Additionally, the user study provided insights into how \projectName{} can integrate into creators' video production workflows, revealing user perceptions of generative video comments.

Collectively, this paper makes the following contributions: \begin{itemize} 
    \item \projectName{}, an interactive system that supports the automatic generation of diverse video comments, enabling creators to receive valuable feedback before publishing their video content. 
    \item A multimodal AI pipeline that integrates multiple data modalities in video—such as visuals, audio, and metadata—along with user personas sampled from a large dataset, to produce diverse, relevant, believable, and helpful video comments for creators. 
    \item A thorough evaluation involving automatic metrics, crowdsourcing, and qualitative user studies, demonstrating \projectName{}'s effectiveness and providing insights for the development of future AI-assisted feedback tools in content creation. 
\end{itemize}

%% file: sections/2_related_works.tex
\section{Related Work}
\label{section:rw}
\projectName{} builds upon prior work on vision-guided natural language generation, feedback tools, and the simulation of human-like behaviors using LLMs.

\subsection{Vision-guided Natural Language Generation}

Natural Language Generation (NLG) focuses on developing machine learning techniques to generate textual content. A subarea of NLG integrates visual input, such as images and videos, to guide the generation process. A prominent example is image captioning, where a model takes an image as input and outputs a textual description of the visual content~\cite{Barnard2016VisionLanguage, Kulkarni2011BabyTalk, kuznetsova-etal-2012-collective, mitchell2012midge, hossain2019ImageCaption}. This has many downstream applications, including accessibility enhancements~\cite{dognin2022image, gurari2020captioning}. Other vision-guided NLG tasks include visual question-answering (VQA)~\cite{antol2015vqa, goyal2017making}, where a model receives an image and a language-based query as inputs and outputs an answer, or visual storytelling, which involves generating coherent narratives based on a sequence of images~\cite{Farhadi2010PictureStory, kim2019glacnetglocalattention, Smilevski_2018, huang-etal-2016-visual, chandu-etal-2019-way}.

Beyond static images, videos serve as another form of visual content, extending imagery with a temporal axis. Accordingly, tasks like video captioning~\cite{song2018deterministic, shen2017weakly, barbu2012videosentences, Tang2021CLIP4Caption, Bogolin2020VNLG, Guadarrama2013Youtube2Text}, audiovisual content recognition~\cite{Tan2011AudioVisual}, and video question-answering (Video QA)~\cite{Zeng2017VideoQA} have been explored. The temporal dimension introduces additional information, which can be excessive for applications requiring concise understanding. Thus, research has been conducted to summarize individual frames into concise video summaries~\cite{chen2024sharegpt4videoimprovingvideounderstanding, wang2024lave, 10.1145/3613904.3642839}. Depending on the video type, different generation tasks arise. For instance, in live-streamed videos, there has been work on generating real-time commentary~\cite{Zeng2021PLVCG, ma2018livebotgeneratinglivevideo, wu2020responselivebotgeneratinglive, duan2020multimodalmatchingtransformerlive}. 

These works are closely related to \projectName{}. However, they primarily focus on generating viewer comments for entertainment purposes and are limited to specific types of videos based on their training data. In contrast, our work aims to generate diverse perspectives intended to assist creators in improving their videos. Additionally, we leverage advanced vision-language models (VLMs)~\cite{openai2023gpt4, chen2024sharegpt4videoimprovingvideounderstanding}, which are more generalizable and capable of handling diverse video inputs and generating varied language outputs, surpassing the capabilities of prior work.

\subsection{Tools for Providing Feedback}

Feedback is essential for improving task performances ~\cite{John2007feedbackpower, kluger1996effects, chinmaye2015Peerstudio}. Traditionally, feedback was gathered through face-to-face and synchronous interactions. However, with the advancement of the Internet, feedback-gathering tools now facilitate asynchronous collection and transfer of rich feedback forms for documentation~\cite{yoon2014richreview}, music~\cite{Colin2024Showandtell}, image editing~\cite{Zhang_2024_CVPR}, and video~\cite{videocrit2016pavel, Ramos2003annotateVideo} through text, audio, pen input, and even gestures. Despite these advancements, such tools still rely heavily on human feedback, resulting in high wait times and significant human effort. Timely feedback is often only possible through peer-based critiques, as seen in MOOC scenarios~\cite{chinmaye2015Peerstudio}.

To address these limitations, automatic feedback tools have emerged, providing real-time, personalized feedback without human intervention across various tasks, such as programming~\cite{Lee2011Programming}, writing~\cite{STEVENSON201451, Nilforoshan_Wu_2018}, and academic assignments~\cite{malik2021generativegradingnearhumanlevel}, helping users iterate and refine their work. Some automated assistance tools for content creators, such as AI Insights by YouTube~\cite{2023madeonyoutube}, suggest textual video outlines based on user input and prior content. Thematic~\cite{thematic2024} analyzes real video comments to generate bulletin-style reports. \projectName{} extends this body of work on automatic feedback tools by contributing a novel interactive system capable of generating video comments from diverse perspectives, providing creators with valuable feedback. It also allows users to interactively explore various generated comments, offering further opportunities for iterative refinement and creative exploration.

\subsection{Simulating Human-Like Behaviors with LLMs}

LLMs, trained on vast datasets and knowledge~\cite{brown2020language, openai2023gpt4}, have demonstrated the ability to perform tasks once thought to be exclusive to humans, such as reasoning ~\cite{wei2022chain} and creative writing ~\cite{dramatron2023, coauthor2022}. Consequently, there has been growing research exploring the potential of using LLMs to simulate human behaviors. One prominent example is Generative Agents \cite{park2023generative}, where LLMs simulate daily life and interactions between virtual characters in a game environment. Prior work has also explored simulating Reddit community comments \cite{park2022social} and synthesizing user study data~\cite{hamalainen2023LLMHCI}. Additionally, numerous studies have employed LLMs to create agents serving various roles, such as educators~\cite{Piro2024MyLearningTalk}, customer service agents~\cite{pandya2023automatingcustomerserviceusing, Saydulu2023Customer}, and even historical figures\footnote{\url{https://www.cnn.com/2023/08/21/tech/khan-academy-ai-tutor/index.html}}. These simulations are often achieved through prompt engineering using a persona or a description of a human background, enabling the models to perform tasks specific to the assigned persona~\cite{jiang2024personallminvestigatingabilitylarge, shao2023characterllmtrainableagentroleplaying, Benharrak_2024, tseng2024talespersonallmssurvey}.  In this work, we build upon this body of research to further explore how LLMs, equipped with personas, can benefit creators by simulating human interactions and feedback on video-sharing platforms, providing insights for content refinement. However, we acknowledge the potential implications of simulating human-like behaviors and ensure that users understand our generated comments are purely simulated, intended as a supplement to real human feedback rather than a replacement.

%% file: sections/3_design_goals.tex
\section{Design Goals}
\label{sec:design-goals}

The objective of our system, \projectName{}, is to automatically generate simulated video comments that provide creators with useful, actionable feedback, while also allowing them to customize and iteratively explore the generated content. To this end, we have established the following design goals (DGs):

\begin{itemize}
    \item \textbf{DG1: Integrating and Leveraging Multimodal Information for Video Comment Generation.} 
    Videos inherently contain rich information across multiple modalities, including visuals, audio, and narration. 
    To comprehensively inform \projectName{}'s comment generation module, we aim to design a pipeline that effectively integrates and leverages multimodal information extracted from videos.

    \item \textbf{DG2: Generating Diverse, Relevant, and Believable Comments.} 
    We aim to design \projectName{} to generate comments that are diverse, relevant, and believable, ensuring they are beneficial to creators.
    We intend to supplement, rather than replace, genuine human feedback by offering an automated and scalable means of generating useful and assorted feedback for creators.

    \item \textbf{DG3: Supporting Users to Interactively Generate Comments.} 
    We plan to design \projectName{} as an interactive tool that engages users in the comment generation pipeline, enabling them to steer the generation according to their preferences, \eg feedback from the targeted audience, and developing in-depth discussions with commenters.     
\end{itemize}

Guided by these design goals, we developed the \projectName{} system as a comprehensive full-stack web application, featuring a frontend user interface (UI) and a backend comment simulation pipeline. 
We first describe the UI and user expereince of using \projectName{} in \Cref{sec:ui_implementation}, followed by the discussion of the backend implementation that empowers the user experience, in \Cref{sec:backend}. We will also highlight how our design goals are tightly connected to specific aspects of the system’s design.

%% file: sections/4_ui_Implementation.tex
\section{\projectName{} User Interface}
\label{sec:ui_implementation}

\begin{figure*}[htb]
    \includegraphics[width=\linewidth] {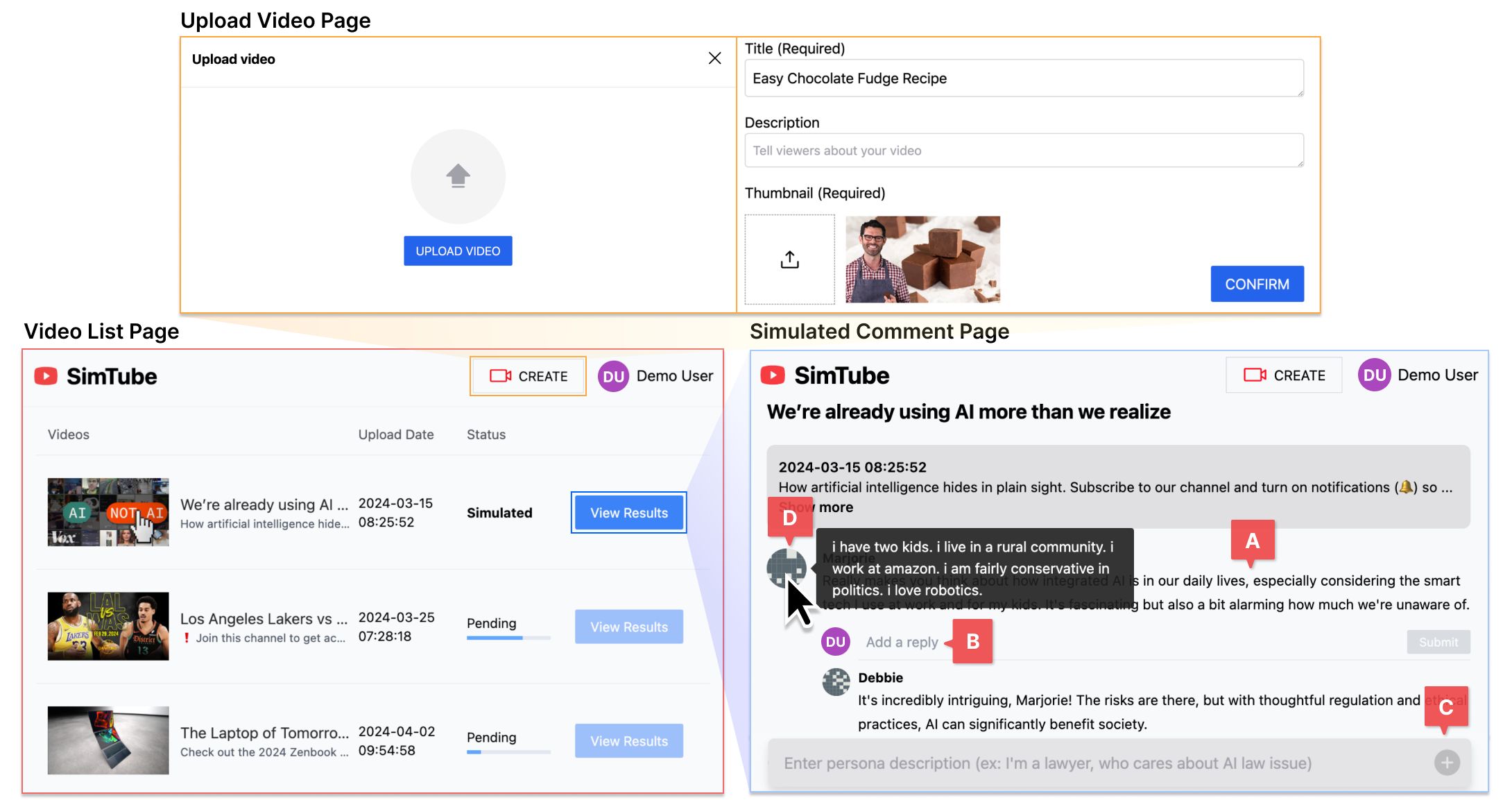}
    \caption{
        Users upload a video via the Upload Video Page, and (A) view the comments generated by SimTube on the Simulated Comment Page, and by (B) replying to a comment or (C) specifying a customized persona, users can ask SimTube to generate new comments. In addition, users can refer to the persona description by (D) hovering on the commenter's icon.
     }
    \label{fig:UI_tags}
    \Description{Caption result demonstration.}
    \vspace{0pt}
\end{figure*}

SimTube's user interface (UI) implements a three-step workflow: \textit{Video Uploading}, \textit{Comment Simulation}, and \textit{Comment Customization}. In the process, users upload a video, after which SimTube's backend pipeline generates related comments. Finally, users can review and interact with the comments. The following subsections will further discuss each component of the workflow.

\subsection{Video Uploading}

The video upload process begins when users click the \textit{Create} button, which opens \textbf{Upload Video} page, a modal interface for video submission, as depicted in the upper section of \Cref{fig:UI_tags}. Users provide essential video metadata, such as title, description, and thumbnail image, similar to what the audience would perceive on popular video streaming platforms like YouTube. The system supports multiple video uploads, with all submissions accessible through the \textbf{Video List} page, as depicted in the lower left quadrant of \Cref{fig:UI_tags}. Upon successful upload, the frontend sends the video along its metadata to the backend pipeline, initiating the comment generation process \textbf{(DG1)}. 
Simultaneously, a progress bar that mirrors the real-time progression of comment generation is incorporated.

\subsection{Comment Simulation}

Upon completion of the simulation process, the \textit{Show Result} button materializes, guiding users to the \textbf{Simulated Comment} page, illustrated in the lower right quadrant of \Cref{fig:UI_tags}. The page displays contextually relevant comments generated by the backend pipeline beneath the video content. Each simulated comment includes comprehensive particulars: the commenter's name, profile picture, and persona — accessible by hovering over the profile pictures, as demonstrated by tag D in \Cref{fig:UI_tags}. Each comment is generated based on different personas, showcasing the diversity of generated comments from different user backgrounds \textbf{(DG2)}. The comment interface deliberately mirrors conventional video-sharing platforms, providing users with a familiar environment for receiving feedback. This design approach ensures that users can focus on evaluating the simulated comments without interface-related distractions, maintaining an intuitive and seamless review experience.

\subsection{Comment Customization}
\label{customize}

\begin{figure*}[htb]
    \includegraphics[width=\linewidth]{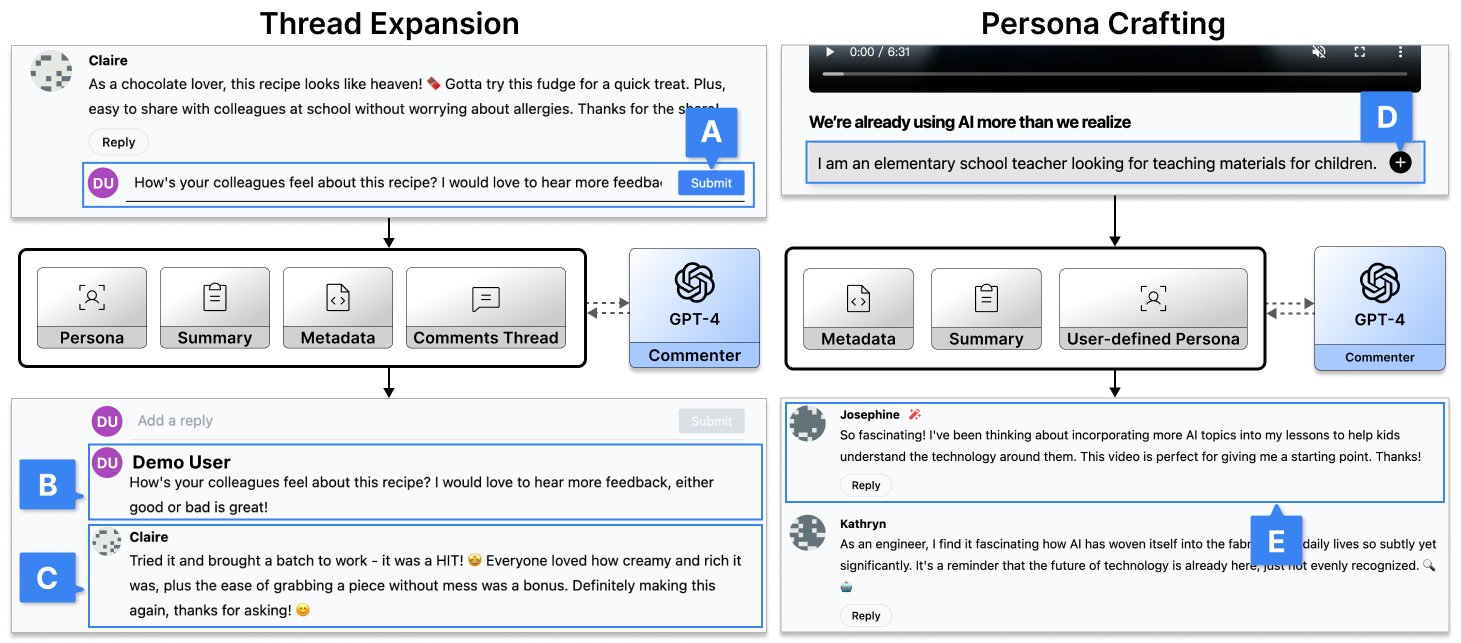}
    \caption{
        \emph{Thread Expansion}: Upon receipt of (A) the user's reply, The thread is expanded by both (B) the user's reply and (C) the generated response of the commenter.
        \emph{Persona Crafting}: (D) Upon user specification of a persona, (E) a new comment is generated in alignment with the video content and the user-defined persona.
    }
  \Description{}
  \vspace{0pt}
  \label{fig:generate_comment_combined}
\end{figure*}

While automatic comment simulation generates diverse perspectives, \projectName{} extends the generation by offering user-directed customization capabilities. This customization framework enables adjustments in comment generation input, allowing users to explore perspectives beyond the initial output \textbf{(DG3)}. The system implements two interaction mechanisms for comment :
\begin{itemize}
    \item \textbf{Thread Expansion}: Users can extend existing discussions by replying to any simulated comment. The dialogue is then deepened with the system generating a follow-up reply with the original persona, as illustrated in the left panel of \Cref{fig:generate_comment_combined}.
    \item \textbf{Persona Crafting}: Users can receive feedback from specific audience's perspectives by defining personas. A new comment will be generated according to the user-defined persona and the video content, as demonstrated in the right panel of \Cref{fig:generate_comment_combined}.
\end{itemize}

%% file: sections/5_backend_implementation.tex
\section{\projectName{} Backend Implementation}
\label{sec:backend}

In the ~\Cref{sec:ui_implementation}, we described the UI of \projectName{} and explained how users could interact with our system to obtain video feedback through generated comments. We now detail our novel generative AI-based backend pipeline that powers these features. Starting from how the audience would perceive a video, from clicking the video that interests them, watching the video, acquiring the content, to leaving their comments. The backend pipeline consists of three primary components, as illustrated in \Cref{fig:system_diagram_example}: (a) a Video Understanding module, which captures the semantics of video content through multimodal summarization; (b) a Persona Query module, which retrieves relevant user personas for providing feedback on the video; and (c) Comment Generation, which combines video understanding and user persona information to generate and present comments in the UI, allowing user interaction.

\subsection{Video Understanding}

The video understanding pipeline digests multimodal content of the video, including the clips, thumbnails, and metadata (\textbf{DG1}), subsequently synthesizing and producing a cohesive textual video summary.
To achieve this within the current models' capability, we decompose the video into visual and auditory components. The pipeline diagram is illustrated in the red block "Video Understanding" tagged (a) in \Cref{fig:system_diagram_example}.

\subsubsection{Audio Transcription}
\label{audio_transcript}

Many videos contain speech dialogue or narration, offering rich information for understanding their themes and key messages. To extract this information, we use Whisper~\cite{whisper} to transcribe the dialogue or narration within the video. Among Whisper’s models, we select the \texttt{Whisper-medium} model for its balanced efficiency and performance. The resulting transcription provides both the dialogue and corresponding timestamps.

\subsubsection{Frame Captioning}

A comprehensive understanding of a video requires integrating crucial visual elements such as objects, scenes, and dynamic information~\cite{Huang_2018_CVPR}. We utilize a powerful vision-language model (VLM), \texttt{LLaVA-NeXT 13B}, to convert visual data into textual descriptions. 
However, most VLMs are designed to process single images, while analyzing video content necessitates capturing temporal continuity across sequential frames. To address this, we propose a unique approach that considers the temporal aspects of video while aligning visual and audio tracks. 
First, we process the visual and auditory tracks simultaneously in chronological order, matching them through timestamps. This frame-dialogue pairing enables the model to better understand the scene and activities with temporal context. 
Additionally, we employ a comic-like assembly of four consecutive visual frames as a single image, referred to as a "panel", which is input into the VLM. We design a VLM prompt to emphasize key visual details, such as objects, scenes, temporal dynamics, and character emotions, as outlined in ~\Cref{app:prompts}. The VLM then processes each panel-dialogue pair, generating a caption for a one-second video segment. This caption, enriched with audio and temporal details, integrates the essential information. Example captions are shown in ~\Cref{app:framecaption}.

\subsubsection{Summary Generation}
\label{video_summary}

Once all video captions are generated, we use an LLM to summarize them along with the transcript \footnote{The audio transcript is used in both the VLM frame captioning and the LLM video summarization, as we found the narration information beneficial for both tasks.}  to create a comprehensive video summary. The captions and transcript are fed into the LLM in chronological order to maintain the temporal flow. However, videos can contain numerous frames, making it difficult to fit all information into a single context window. To overcome this, we utilize \texttt{Claude 1.6}, an LLM with an extended context window of 200K tokens. Based on our testing, this capacity is sufficient to accommodate all frame captions and the audio transcript of a 25-minute video. The LLM generates a video summary and extracts keywords, which are then used in the next stage—persona querying.

\begin{figure*}[htb]
\includegraphics[width=\linewidth]{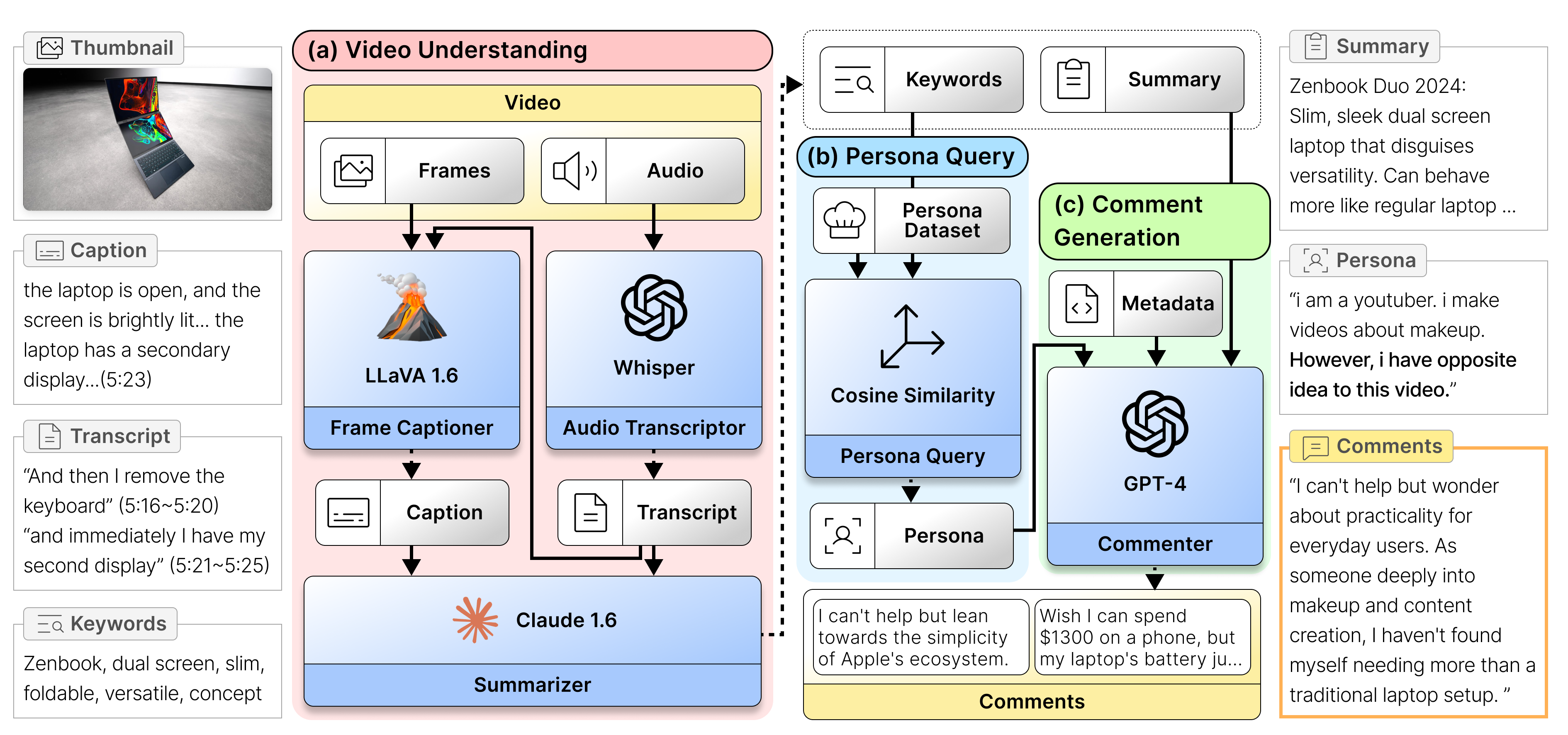}
     \caption{
        \projectName{} leverage VLM and \textbf{Whisper} to process videos in visual and audio components per modality. Personas are then queried based on the video content for comment generation. 
        The system diagrams, situated centrally in the figure, along with the intermediate outputs on both flanks, exemplify this process.
     }
    \Description{Caption result demonstration.}
    \label{fig:system_diagram_example}
\end{figure*}

\subsection{Persona Query}
\label{query_persona}

In our initial testing, generating comments directly from the video summary obtained in the previous stage produced informative and helpful comments; however, they often lacked sufficient diversity in opinions and variety. Inspired by prior work leveraging user personas to simulate diverse generation \cite{personachat2022, jiang2024personallminvestigatingabilitylarge}, we incorporated persona information into our generation pipeline by developing the Persona Query system. This system queries multiple related personas for comment generation, as shown in the blue block labeled (b) Persona Query in \Cref{fig:system_diagram_example}.

\subsubsection{Persona Dataset}

A persona is a textual description of a person's personality, background, and hobbies, and other demographic information. We include personas in our system to help the LLM generate diverse outputs~\cite{jiang2024personallminvestigatingabilitylarge} (\textbf{DG2}). Specifically, we use \textsc{PersonaChat}~\cite{personachat2022}, a popular dataset containing over 8,000 personas, to enhance the diversity of user backgrounds in our comment generation pipeline.

\subsubsection{Querying Persona}

In our preliminary testing, irrelevant personas usually lead to low-quality comments; therefore, we focus on those most relevant to the video content.
To achieve this, we measure the similarity between personas and the video keywords obtained in \cref{video_summary}.
Compared to the whole video summary, keywords concisely cover the video topic in just a few representative words, making the process faster and more accurate.
We first embed all personas and keywords into text embedding using the \texttt{text-embedding-3-small} model from OpenAI. After that, we calcualte the cosine similarity between each persona and the keywords. We then rank the personas based on their similarity to the video’s keywords, selecting the top 30 most relevant personas. Once the most relevant personas are retrieved, the pipeline uses them in combination with other data, including the video summary and metadata, to generate video comments.
    
\subsection{Comment Generation}

Our system supports four types of simulated comments using data from earlier stages. The first two types are automatically generated by the system, while the last two are user-initiated and customizable, offering quick and interactive feedback \textbf{(DG3)}: 
\begin{enumerate}
    \item \textbf{Primary Comments}: The initial comments that appear directly under the video.
    \item \textbf{Thread Comments}: The generated responses to existing comments, structured hierarchically as threaded replies beneath primary comments.
    \item \textbf{Response Comments}: The comments generated for responding to user's reply, contributing to a discussion thread under the replied comment.
    \item \textbf{Custom Persona Comments}: The comments generated based on a user-defined persona, tailored to user specifications.
\end{enumerate}

After the video is pre-processed by the upstream components, our system directly generates 30 comments, consisting of 70\% primary comments and 30\% thread comments. Additional generations can be triggered to obtain more comments if needed. Below, we discuss how each type of comment is generated.

\subsubsection{Primary Comments}

We prompt the LLM with a comment generation instruction, queried personas, and video metadata obtained from the earlier stages of backend processing. We design a role instruction that guides the model to emulate typical YouTube commenting behavior through few-shot prompting \cite{wang2020generalizing}, using real YouTube comments as examples. Subsequently, we incorporate the retrieved personas from \cref{query_persona} to ensure diversity in the generated comments. The video metadata includes details such as the video title, description, author, and thumbnail description, along with the summary and keywords generated previously, encapsulating information a viewer might perceive when watching a YouTube video. Once the comments are generated, a random name and icon are assigned for authenticity. Names are selected from an SSA dataset~\cite{2023USAName} containing over 10,000 common U.S. baby names, while icons are uniquely generated using \textit{Boring-avatars}~\cite{2023BoringAvatar}.

\subsubsection{Thread Comments}

To simulate discussion threads, the system utilizes different personas to respond to primary comments, adding depth to the conversation and mimicking the extended exchanges commonly seen on video-sharing platforms where viewers engage in discussions. This is achieved by randomly selecting a portion of the primary comments and performing another LLM inference using a prompt that includes all relevant video data (as used for primary comment generation), the selected primary comment, and a new persona. This process generates the thread comments. An example of thread comments is shown in the right panel of \cref{fig:UI_tags}.

\subsubsection{Reply Comments}

Our tool enables users to interact with pre-generated comments (Thread Expansion feature detailed in \cref{customize}), similar to engaging with other people in an online forum. When a user replies to a comment, the system generates a relevant response, providing immediate feedback. The method for generating these responses is identical to that of Thread Comment, with the distinction that the user's reply and the replied comment are included. The system processes these comments within approximately 10 seconds. An example of reply comments is labeled by tag C in \cref{fig:generate_comment_combined}.

\subsubsection{Custom Persona Comments}
The system also supports users in customizing personas for further exploration and experimentation. When a user-defined persona is provided, the system generates comments using the same approach as for primary comments, but tailored to the user’s specified persona. For inspiration on persona customization, users can hover over an existing comment's avatar to view example persona descriptions from other comments (as shown by tag D in \Cref{fig:UI_tags}). This type of comment is generated when users utilize the Persona Crafting feature described in \cref{customize}.

These four types of comments are presented in the frontend UI for users to read, engage with, and provide feedback on their videos. In the following two sections, we will discuss comprehensive studies on the quality of the generated comments, as well as user perceptions of our system and its comments, and how it might integrate into their video creation workflow.

%% file: sections/6_evaluation.tex
\section{Quantitative Evaluation}

We conducted extensive quantitative evaluations verifying the ability of \projectName{} to produce diverse and relevant comments.
Specifically, we adopt both crowd-sourced, human-annotated evaluations as presented in \Cref{sec:cs_eval}, and automatic metrics widely used for natural language generation tasks, such as Self-BLEU, Distinct-Ngram, BERTScore, etc., as presented in \Cref{sec:auto_eval}.

\subsection{Crowd-Sourced Evaluation}
\label{sec:cs_eval}

We conducted a crowd-sourced evaluation to assess the effectiveness of \projectName{} from human perspectives. Given the widely accepted understanding that LLMs can generate sentences with human-level fluency~\cite{bubeck2023sparksartificialgeneralintelligence}, our study focuses on three high-level aspects: \textit{Relevance}, \textit{Believability}, and \textit{Helpfulness}, to ensure alignment with our design goal \textbf{(DG2)}.

\begin{itemize}
    \item \textit{Relevance}: refers to the degree to which comments are aligned with the video's content. 
    \item \textit{Believability}: indicates the likelihood that users would consider the comment credible. 
    \item \textit{Helpfulness}: assesses the extent to which comments can provide valuable feedback to content creators, aiding in the enhancement of their videos and potentially inspiring new ideas.
\end{itemize}

\subsubsection{Videos and Comments}

We selected eight videos from YouTube, each covering one distinct video genre listed in \Cref{app:videoCategories}.
For each video, we sampled 30 comments from three categories:
\begin{enumerate}
    \item \textbf{Real Comments} are from the real YouTube users commenting on this video. We randomly sampled 30 comments from the top 1000 popular comments on each video, ranked by YouTube based on their meta-ratings, to exclude any offensive or inappropriate remarks~\cite{Siersdorfer2010predictcomment}.    We also filtered out non-English comments to exclude the effect of different languages.
    \item \textbf{Persona-based Comments} were generated by the full \projectName{} system, including Primary Comments and Thread Comments.
    \item \textbf{No Persona Comments} were generated by an ablated version of \projectName{} without the persona components.
\end{enumerate}

\subsubsection{Survey Protocol}

We recruited 25 participants on Upwork to evaluate comments according to the three aspects, utilizing a 7-Point Likert Scale.
We collect the ratings from the participants using Google Forms. 
Each form includes one YouTube video link, five video quizzes, and three evaluation sections with different criteria for the same 30 comments.
To ensure the quality of responses, we structured the forms into five stages: 1) Watch the Video, 2) Video Quiz (see \Cref{app:FormQuestion}), 3) Video Summary, 4) Comments Evaluation, and 5) Form Feedback.
A validated reply must achieve an 80\% correction rate in the quiz in stage 2 and provide an overall correct video summary in stage 3.
The video summaries collected in stage 3 are reused as the ground truth of human-like detailed video summaries for later use.

\subsubsection{Results}

\begin{figure}[htb]
  \includegraphics[width=\columnwidth]{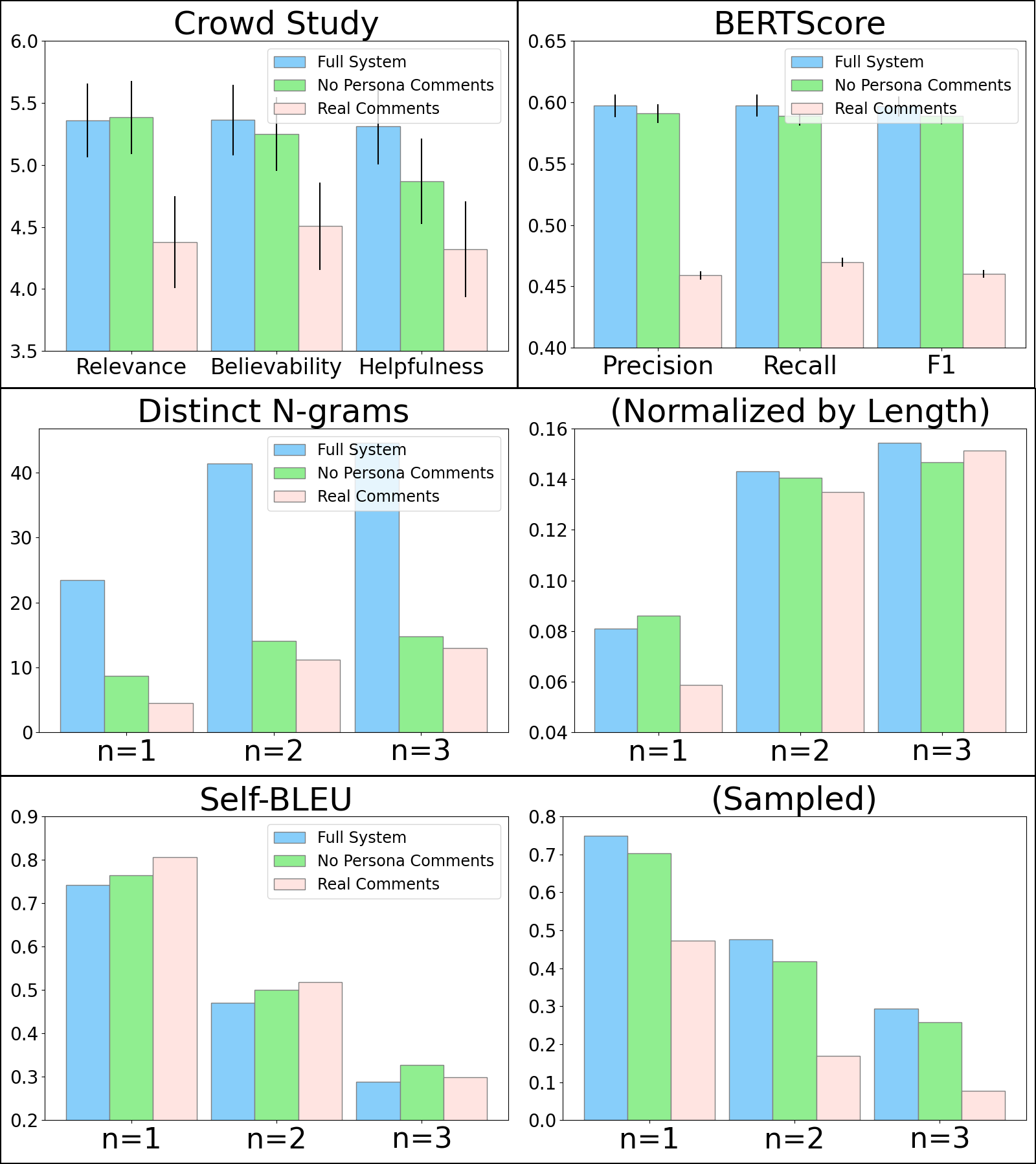}
    \caption{
        This plot includes quantitative results from Crowd-Sourced Study and other automatic metrics measuring diversity such as BERTScore, Distinct N-grams, and Self-BLEU.
     }
    \label{fig:eval_diversity}
    \Description{}
\end{figure}

The results from our crowd-sourced study, displayed in the "Crowd Study" plot in \Cref{fig:eval_diversity}, show significant differences between the two configurations of \projectName{}-generated comments and real comments. Specifically, SimTube comments with the persona configuration were rated significantly higher in \textit{relevance}, \textit{believability}, and \textit{helpfulness} compared to real comments (p < 0.05). Similarly, comments generated without a persona (No-persona) also outperformed real comments in \textit{relevance}, \textit{believability} (p < 0.05). However, the comparison between the full system and the No-persona configuration did not yield a statistically significant difference.

We used the Wilcoxon Signed-Rank Test for paired comparisons, with Bonferroni Correction applied to control for multiple comparisons across these three conditions. These results highlight the effectiveness of SimTube-generated comments, with or without personas, in delivering feedback perceived as more valuable than that found in real comments. 

While sampling candidates from popular real comments effectively eliminate most nonsensical comments, it is noteworthy that these popular comments are often brief and offer limited informative value for creators. In some instances, they are purely laudatory, offering no substantial critique or feedback. Conversely, with LLM's outstanding writing skills and inclusive attitude, generated comments usually stimulate broader discussion and a multitude of potential insights. This result aligns with recent studies showing that LLM output can outperform human-written ones~\cite{Goyal2022NewsSA}.
Some comments and their average scores of this evaluation are shown in the \Cref{app:SampledRealComments}.
The raw average scores and lengths of comments can be found in \Cref{app:CrowdStudyResult}.

\subsection{Automatic Metrics}
\label{sec:auto_eval}

Besides the crowd-sourced evaluation, we also conducted extensive evaluations using automatic metrics~\cite{li-etal-2016-diversity, zhu2018texygen, lin2004rouge, bert-score, lin-chen-2023-llm} widely adopted in natural language generation tasks~\cite{liu2023g, yasunaga2022linkbert, gong2022diffuseq, lu2022learn}. Similar to the crowd evaluation, we assess three types of comments: \textit{Full System (with Persona)}, \textit{No Persona}, and \textit{Real Comments}. For the automatic metrics, we focus on two widely evaluated aspects in NLG: 1) Diversity and 2) Relevance of comments. To minimize variance in the evaluations, we generated a larger number of comments for the automatic metrics assessment. The total number of comments for \textit{Full System}, \textit{No Persona}, and \textit{Real Comments} are $671, 1141, 18338$, respectively, while the average lengths of the comments are $286.18, 100.1, 85.58$ characters.

\subsubsection{Diversity}

A diverse set of comments offers varied perspectives, encompassing different word choices and comment topics. 
Typically, generated comments, particularly those from the Full System (persona-based), exhibit the highest diversity in wording, while Real Comments display greater diversity at the semantic level. 
We also noted that a subset of Real Comments appeared to be highly repetitive, as indicated by the significant variation in the sampled size of the Self-BLEU Score.

\myparagraph{Distinct N-grams}
A greater count of distinct N-grams indicates a higher level of word-level diversity.
Generated comments, especially the Full System (Persona-based) comments, achieve the highest average number of distinct n-grams per comment, as depicted in the plot titled "Distinct N-gram" in \Cref{fig:eval_diversity}.
Considering the length variations across the conditions—with Real Comments generally being short~\cite{thelwall2012commenting}—the right side of the plot titled "Distinct N-gram" in \Cref{fig:eval_diversity} presents length-normalized results.

\myparagraph{Self\-BLEU}
Self-BLEU inspects the similarity of intra-group comments. 
A higher Self-BLEU score implies a higher similarity among a group of comments and lower diversity.
The right side of the plot titled "Self-BLEU" in \Cref{fig:eval_diversity} demonstrates the Self-BLEU scores of same number of comments across different sources.
The left half of the plot indicate the Self BLEU score of a larger group number of comments (num=$671,1141,18338$). We noticed that real comments have a higher score as the sample number increases, which may imply that a part of real comments are duplicated.

\myparagraph{BERTScore}
Intra-group BERTScore checks the semantic similarity in a group of comments. A higher intra-group BERTScore indicates a lower diversity.
The real comments score the highest semantic level diversity, as shown in the plot title "BERTScore" in \Cref{fig:eval_diversity}.

\subsubsection{Relevance}

Some Real Comments can be off-topic and divergent, causing confusion and high semantic diversity scores.
A relevant comment is on-topic and pertinent to the video content.
We measured the word and semantic similarity between comments and video content. 
The ground-truth video summaries were compiled from 30 User Study participants using Claude 3.5 Sonnet.
Generally, our generated comments were more relevant to the video on both levels.

\begin{figure}[htb]
  \includegraphics[width=\columnwidth]{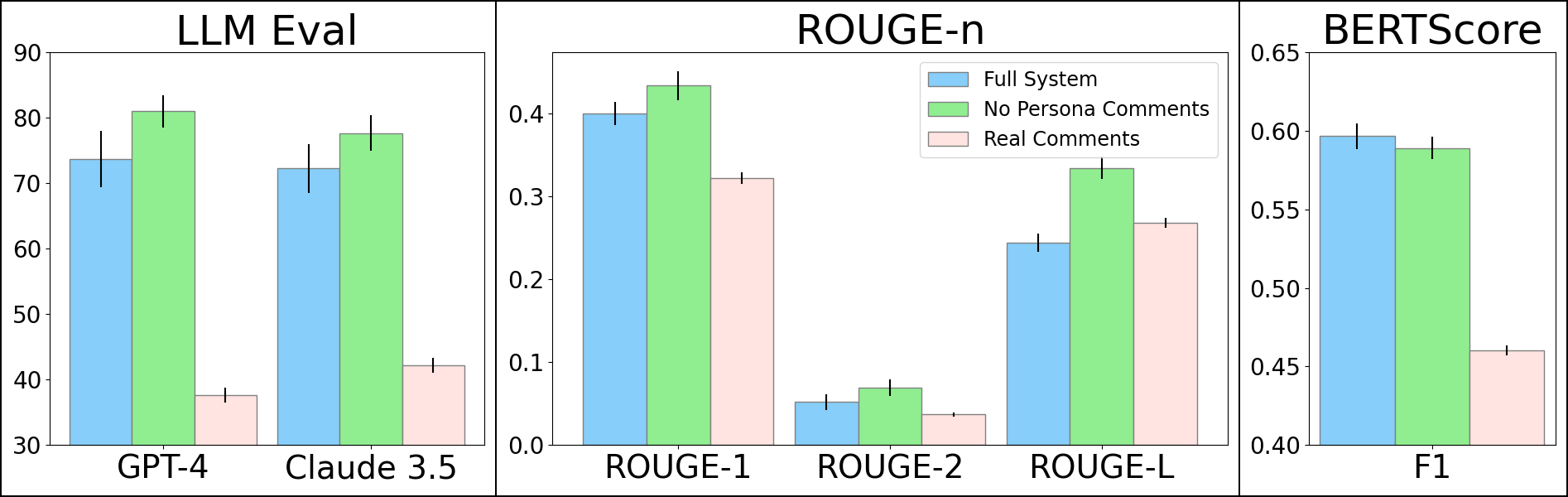}
    \caption{
        This figure contains quantitative results from automatic metrics measuring relevance between targeted text and video content, such as LLM Eval, ROUGE-n (n=1, 2, L), and BERTScore.
     }
    \label{fig:eval_relevance}
    \Description{}
\end{figure}

\myparagraph{LLM Evaluation}
We adopted LLM evaluation, which has shown high correlation coefficients with human ratings among commonly used automatic metrics~\cite{lin-chen-2023-llm}. To mitigate potential bias from LLMs towards their own generated comments~\cite{panickssery2024llmevaluatorsrecognizefavor}, we utilized two different models, \textbf{Claude 3.5 Sonnet} and \textbf{GPT-4}, to assess both real comments and those generated by our system. The LLMs evaluated each target comment against the video summary and rated its relevance on a scale from 0 to 100, where a higher score indicates greater relevance. The results show that the generated comments significantly outperform the real comments, demonstrating closer relevance to the video, as illustrated in the plot titled "LLM Eval \Cref{fig:eval_relevance}. 

\myparagraph{ROUGE}
ROUGE is a recall-based metric that evaluates the relevance of a comment at the word level. 
The plot titled "ROUGE-n" in \Cref{fig:eval_relevance} presents the precision score, which is the proportion of words in the comment that are also found in the video summary. Mostly, generated comments exhibit greater relevance to the video across all ROUGE metrics.

\myparagraph{BERTScore}
BERTScore evaluates the semantic-level relevance of embeddings between comments and the given video summary.
Hence, a higher BERTScore indicates greater comment relevance to the video.
To compute the BERTScore, we utilize the model \texttt{microsoft/deberta-xlarge-mnli} from Hugging Face, which has the highest correlations with human evaluation currently in the WMT16 benchmark.
As shown in the plot titled "BERTScore" in \Cref{fig:eval_relevance}, The generated comments outperform real comments and demonstrate a statistically significant difference in semantic-level relevance.

\subsection{Conclusion}

According to the crowd-sourced evaluation and the evaluations using various automatic,
the comments generated by our system display superior word-level diversity, while Real Comments showcase better semantic diversity.
Although a few real comments cover distinct common topics, clusters of real comments may be highly similar, as reflected by the Self-BLEU score.
Concerning relevance to video content, generated comments outperform Real Comments in word-level, semantic-level, and LLM evaluations.
In comparison to Real Comments, generated comments tend to be more on-topic, authentic, and differentiated, offering a potent source of inspiration.
Despite their limited semantic diversity, the scalability, rapid production, and pre-publication availability make generated comments an advantageous preliminary source of inspiration and feedback complementing Real Comments, particularly before the formal publication of videos.

%% file: sections/7_user_study.tex
\section{Qualitative User Study}
\label{sec:user_study} 
In addition to the quantitative evaluation, we conducted a qualitative user study to gather feedback on the comments generated by \projectName{} and to assess the system's utility. We also aimed to understand creators' perspectives on how \projectName{} could integrate into the broader video production workflow.

\subsection{Participants}
We recruited eight experienced content creators, each with a minimum of one year of video filming and editing experience (M = 5.06 years, SD = 4.76 years, range: 1–13 years). Two participants were full-time creators. The age range was 22 to 33 years (M = 26.38 years, SD = 4.27 years), and the gender distribution included five males and three females. Participants were recruited through a university course forum, a Reddit video editing community, and direct outreach on social media. The group represented a variety of video genres, including travel vlogs (P1, P2, P7), airline reviews (P3), Instagram reels (P5), street interviews (P6), tech news (P4), and tech reviews (P8).

\subsection{Study Protocol}
Each study session lasted between 1 to 1.5 hours. Before the session, participants were asked to prepare either a finalized or rough-cut video, including a thumbnail, title, and description. They could optionally provide up to three videos. These videos served as material for comment simulation, enabling participants to evaluate \projectName{}'s capabilities and limitations. At the beginning of each session, participants were briefed on the study protocol and completed a pre-study questionnaire. They then interacted with the system to explore and assess the comments generated by \projectName{}. After using the system, we conducted semi-structured interviews to gather insights into their perceptions of both the tool and the quality of the generated comments.

\subsection{Findings}
We now present the key findings identified in the interviews.

\subsubsection{Creator's Perception of Generated Comments}
Participants generally expressed a positive attitude toward the comments generated by \projectName{}. P1 appreciated how the system provided insights into specific audience viewpoints. Similarly, P2 emphasized the value of diverse perspectives offered by the system-generated comments. P6 highlighted the speed and cost-efficiency of the generated comments, expressing interest in uploading multiple video versions for comparison: \textit{"I might want to upload every version of my video and make an intra-video comparison to further observe the differences between versions."} P4 suggested that generated comments could serve as supplementary feedback alongside real comments but still preferred genuine feedback from human audiences or friends at this stage, as his target audience is human rather than AI.

Our system provides an advantage over traditional methods of gathering extensive feedback, allowing creators to receive input without disclosing unpublished videos. P6, a novice YouTuber, stated, \textit{"I usually need feedback from my friend (before uploading my video)... (Generated comments) are like having more friends to share my new video clips with and get their opinions."} Participants also commended the scalability of \projectName{}. For example, P4, a YouTuber with over 460,000 subscribers, mentioned they only received 126 comments on his latest video in two days, whereas \projectName{} can generate over 5,000 comments for a 20-minute video within a single day. The system also enables users to create in-depth and interactive discussions quickly, whether by responding to comments or assigning personas to the system. Lastly, unlike real comments, LLM-generated text tends to be more moderated, helping avoid cyberbullying and reducing negative effects on users. P4 noted that real hateful comments could be harmful to content creators. \textit{"Generated comments, while generally free from harshness, still present opposing ideas, which is helpful for content creators."} (P4)

\subsubsection{Persona's Effectiveness on Comment Generation}
Persona is a key component we integrated into \projectName{} to support diverse feedback generation, and we aimed to understand its impact on users' perceptions. 
We observed mixed reactions: some participants found persona-based comments engaging and relevant (P2, P4, P6), while others viewed them as unrealistic or unnecessary (P1, P5, P8). In contrast, no persona comments were concise and video-focused (P2, P5, P7), but they sometimes lacked inspiration (P4), detail (P6), and engagement (P4). P4 stated, \textit{"Persona-based comments look informative, just like an audience resonating strongly with my video."} On the other hand, P2 commented, \textit{"Their replies don't bring new ideas."} We also observed that some persona-based comments contained irrelevant information because the sampled personas were less relevant to the video, which could cause confusion. Overall, we found that while persona integration adds depth and detail to comments and can inspire users, it may also introduce inconsistencies.

\subsubsection{Threads Expansion and Persona Crafting}
\projectName{} allows users to interactively engage in conversations through threaded replies and persona customization. Participants found the threaded reply feature useful, benefiting from the opportunity to engage in multiple rounds of interaction. For example, P1 initially had trouble using feedback to improve her video, but after replying to ask for specific suggestions, she gained valuable insights that helped her make improvements. Similarly, P3 received a comment on her budget airline review video, suggesting she try a full-service airline. She responded by asking the AI commenter to compare their experiences, which led to a detailed reply and the idea for a future comparison video. P6 also asked for advice on "broadening horizons" in her video content, and the system provided her with detailed, personalized suggestions that matched her video’s theme.

The persona customization feature enables users to generate more targeted feedback using personas that may not have been included in our dataset. For example, P7 applied personas such as a mom, critic, and advisor for her self-reflective video, finding that these tailored personas provided feedback that felt realistic and aligned with her needs. Similarly, P1 used the persona of an English-speaking East Asia travel enthusiast. She felt the generated comment accurately captured her target audience's perspective: \textit{"Wow, this vlog seriously makes me miss Seoul. Your adventures at Ewha and the nightlife clips brought back so many memories! Can't wait to see what Japan brings."} 

\subsubsection{Integrating SimTube into Video Production}
Participants identified multiple stages in their video production workflows where SimTube could be effectively integrated, aligning closely with professional film production processes such as topic brainstorming, outlining, footage capture, editing, and finalization~\cite{owens2023}. Many participants (P4, P6, P8) emphasized creating rough-cut versions or teasers during the editing phase, enabling collaboration and feedback from sponsors or team members. As P4 explained, \textit{"A rough-cut lets my team and sponsors give feedback before we proceed further"}. This suggests that SimTube could enhance this process by providing automatic, diverse feedback on uploaded rough cuts. P6 added, \textit{"I can seamlessly integrate SimTube into my workflow and collect more feedback with minimal effort by uploading the rough-cut version or any segments whenever I complete one."}

Beyond assisting with existing content, SimTube could also inspire new video topics. Participants (P1, P3, P6) highlighted that AI-generated comments led them to explore new ideas. For instance, P1, a travel vlogger, received recommendations for famous tourist spots like Ikseon-dong Hanok Village after uploading a Korean vlog, even though these places were not featured in the video. \textit{"SimTube correctly listed all my itineraries based on my narration, which helped me plan new vlogs,"} P1 noted, demonstrating the system’s ability to generate contextually relevant insights.

SimTube could also influence ongoing video production. P6 uploaded a half-finished street interview video on student lifestyles, and the persona-based comments generated by SimTube extended the discussion between the host and interviewee, introducing new topics such as time management for university students. \textit{"It prompted me to explore this theme further and enriched my video,"} P6 shared, illustrating how SimTube’s integration can guide and enhance content creation throughout different stages of the workflow.

%% file: sections/8_limitations.tex
\section{Discussions and Future Work}

We have presented the design, implementation, and evaluation of \projectName{}. In this section, we discuss the key findings from our study, the limitations of our approach, and potential areas for future development.

\subsection{Expanding \projectName{}'s Pipeline}
\projectName{} advances automated feedback tools within the video creation workflow by generating pre-publication comments that enable creators to reflect on and refine their content. However, our current focus is limited to audio/visual semantics, addressing only a portion of the video production. Key elements such as visual effects, audio enhancements, and editing techniques—crucial for maximizing viewer engagement—are not yet incorporated. Additionally, while \projectName{} can generate comments for general video content, it currently does not consider inherent variations in video like genre, style, or cultural context. These areas present opportunities to expand \projectName{}'s computational pipeline to accommodate additional contextual information and enable more customized comment generation. Future improvements could also include handling longer video inputs and enhancing the overall quality of language generation to provide more nuanced and useful feedback.

\subsection{Integration into Video Production Workflows}

While we have assessed the quality of our generated comments, the system has yet to be deployed in real-world settings. Future research should explore integrating \projectName{} into video editing tools \cite{wang2024lavellmpoweredagentassistance, huh2023xiang} or production environments to evaluate its overall impact. Qualitative studies could further investigate how the system complements professional workflows, providing deeper insights into its practical utility. Relevantly, the system  currently generates comments only based on a single version of the video. However, creators often produce multiple iterations to determine the best result. By analyzing a series of video edits, the system could generate comparative feedback that highlights differences between the current and previous versions, enabling users to refine their work more effectively by leveraging the strengths of each iteration. Expanding to handle multiple versions, however, introduces challenges related to system scalability and processing efficiency that warrant future explorations.
 
\subsection{Improving Helpfulness of Comments}
Our automated evaluations indicate that LLM-generated comments sometimes surpass online user feedback in several metrics, including helpfulness. This may be attributed to the nature of online comments, which often lean toward opinionated or humorous remarks rather than constructive feedback. In contrast, our system’s comments are specifically designed to provide actionable insights. A potential future direction involves integrating professional feedback data into the pipeline, refining models with expert insights to enhance guidance quality. Nonetheless, general audience feedback remains valuable; our study demonstrates that even non-expert input can inspire creators and positively impact their creative processes.

\subsection{Implications of AI-Generated Comments}
Lastly, we recognize the implications of using AI for human-like comment generation, including potential unintended outputs and misuse. Our goal is to manage these risks responsibly, ensuring that tools like \projectName{} remain focused on supporting and enhancing creators' work. For instance, P4 in our study observed that the system was able to generate critical yet constructive comments in language that is ``\textit{generally free from harshness}'', in contrast to real comments that sometimes contain offensive language and negatively impact creators. Moving forward, we plan to advance \projectName{} with ethical considerations at the forefront, maintaining its supportive role in the creative process.

%% file: sections/9_conclusion.tex
\section{Conclusion}
We have presented \projectName{}, a novel system that automatically generates video comments, offering creators valuable feedback to enhance their videos before publishing. We detailed the design and implementation of our system, focusing on the comment generation pipeline and the user interface. The pipeline utilizes advanced generative AI models to process multimodal video data and incorporate user personas, resulting in diverse and tailored comments. Our user interface supports user-steerable comment generation, allowing users to influence the output based on their responses and defined personas. We conducted a comprehensive evaluation using both quantitative and qualitative methods, including human crowdsourced assessments, automated metrics, and a user study with experienced content creators. We hope our work establishes a foundation for future development of automatic feedback and critique tools in content creation systems.

%% file: sections/10_Appendix.tex
\section*{Appendix}

\section{YouTube Video Category List}
\label{app:videoCategories}
\begin{enumerate}
    \item Animation and Film
    \item Autos and Vehicles
    \item Music Videos
    \item Pets and Animals
    \item Sports
    \item Travel \& Events
    \item Gaming
    \item Comedy
    \item People and Blogs
    \item Entertainment
    \item News and Politics
    \item How to And
    \item Education
    \item Science and Technology
    \item Non Profit \& Activism
\end{enumerate}

\section{Prompts}
\label{app:prompts}
\begin{quote}
    {\small
    \texttt{[Generate Frame Caption] \\
    You are an AI visual assistant that can generate audio description for a video clip. You receive a image of 4 frames, which are sampled during 4 seconds of the video clip. You also receive the audio caption during the 4 seconds. \\ \\
    The 4-th frame is the current frame, using the provided frames and audio caption, generate an audio description of the current frame in a detailed manner. Include details like object counts, position of the objects, relative position between the objects, the visual composition, the emotion, what might be going on during the clip, etc. Imagine describing the video clip to someone who cannot see the clip. 
    }
    }
\end{quote}

\section{Sampled Frame Captions}
\label{app:framecaption}
Pasting four frames in one image and corresponding narration help VLM identify temporal information, the activity, and the scene.
\begin{figure}[htb]
  \includegraphics[width=\columnwidth]{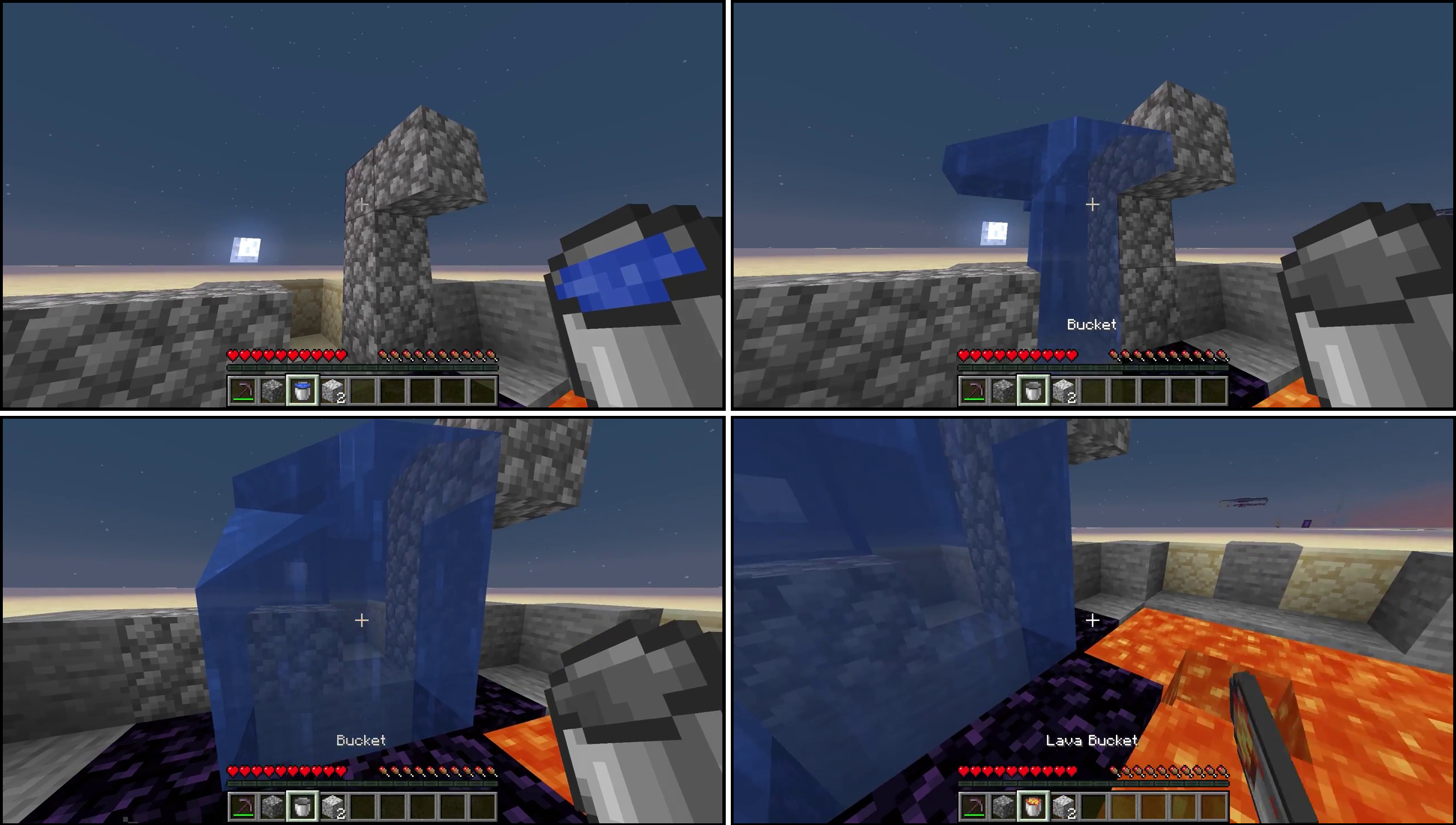}
  \caption{The frames are captured in a \href{https://www.youtube.com/watch?v=dYO9_MoU4Vw}{gaming YouTube video}\protect\footnotemark \ between 0:45 and 0:48.}
  \Description{}
\end{figure}
\footnotetext{\url{https://www.YouTube.com/watch?v=dYO9_MoU4Vw}}
\begin{quote}
    \textbf{Frame Caption}: In the current frame, a player is seen holding a blue bucket, standing in front of a large, glowing lava pit. The player appears to be \ul{in the process of filling the bucket with lava}, as \ul{indicated by the audio caption}. The lava pit is located in the center of the image, with the player standing to its right. The player's position and the direction of their gaze suggest they are focused on the task at hand. The overall scene suggests a moment of gameplay in a video game, possibly involving resource gathering or crafting.
\end{quote}

\begin{figure}[htb]
  \includegraphics[width=\columnwidth]{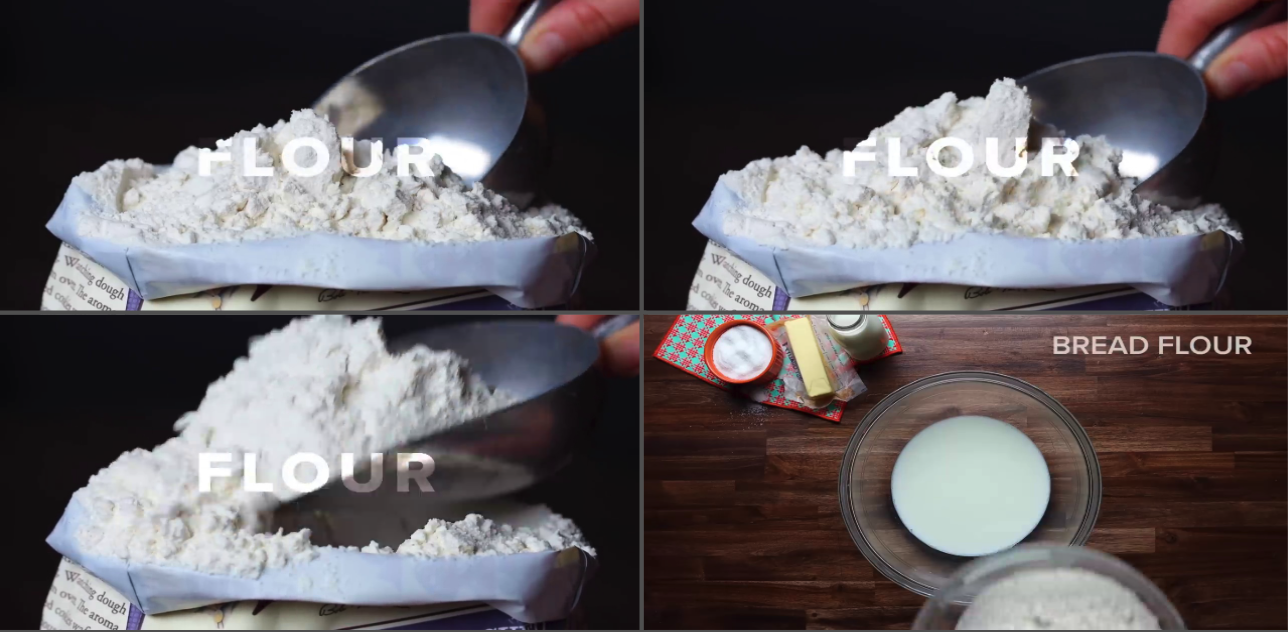}
    \caption{The frames are captured in a \href{https://www.YouTube.com/watch?v=djnNkLi_K6E}{cooking YouTube video}\protect\footnotemark \ between 0:57 and 1:00.}
  \Description{}
\end{figure}
\footnotetext{\url{https://www.YouTube.com/watch?v=djnNkLi_K6E}}
\begin{quote}
    \textbf{Frame Caption}: A hand is seen holding a silver spoon, which is \ul{scooping white flour from a container}. the container appears to be a plastic bag, and the flour is being poured into a glass bowl. the bowl is placed on a wooden surface, and there are other ingredients nearby, including a bottle and a jar. the scene suggests that someone is in the process of preparing a recipe that requires a significant amount of flour. the action of scooping the flour indicates that the person is carefully measuring the ingredients for the recipe. the overall composition of the image suggests a home cooking environment.
\end{quote}

\section{Sampled Form Questions}
\label{app:FormQuestion}
\begin{quote}
    \small{
    What did they use to send the garlic bread to the sky? \\
    (A) Weather balloon \\
    (B) Hot air balloon \\
    (C) Helium balloon \\
    (D) A drone
    }
\end{quote}

\begin{quote}
    \small{
    How did the host end the video? \\
    (A) Express his love \\
    (B) Take a photo with the crowd \\
    (C) Talk in his studio \\
    (D) Pet a puppy
    }
\end{quote}

\begin{quote}
    \small{
    What is the first method introduced? \\
    (A) Baked Chicken \\
    (B) Roasted Chicken \\
    (C) Boiled Chicken \\
    (D) Broiled Chicken
    }
\end{quote}

\section{Comment Length to Rating}

\label{app:CrowdStudyResult}
\begin{table}[H]
    \centering
    \caption{Average Length and Score for Relevance, Believability, and Helpfulness Across Conditions}
    \begin{tabularx}{\columnwidth}{l*{3}{X}}  
        \toprule
        Group          & Full System & No Persona & Real \\
        \midrule
        Average Length & 233.4625 & \textbf{97.8875}  & \ul{107.7625}          \\
        \midrule
        Relevance      & \ul{5.248} & \textbf{5.313} & 4.295           \\
        Believability  & \textbf{5.293} & \ul{5.181} & 4.401          \\
        Helpfulness    & \textbf{5.213} & \ul{4.719} & 4.194           \\
        \bottomrule
    \end{tabularx}
\end{table}

\section{Confusing Persona-based Comment Examples}
\label{app:ConfusingComments}
Related content is underlined and confusing comments is in italic.
\begin{quote}
    \textbf{Video Title}: Grad School Visiting and Bar Exploration in NYC.
    
    \textbf{Persona}: "i like to go to the farmer s market every weekend. i am committed to the environment. i live in nashville. i play music in my spare time. \ul{i work in academia at a local university}. However, i have opposite idea to this video. Will leave different comment with different perspective."
    
    \textbf{Comment}: "Love the diversity of experiences you shared from the academic to the social, \textit{but let's also chat about sustainable travel and how we can enjoy cities responsibly. Greetings from another academic here in Nashville. Let's keep pushing for a greener earth while we explore!}"
\end{quote}

\begin{quote}
    \textbf{Video Title}: Review: Galaxy S24 Ultra - Galaxy AI Comprehensive Test and Smartphones Comparison.
    
    \textbf{Persona}: " i really enjoy reading the news. \ul{i love new technology}. \textit{i try to find easy ways to make money. i wish i could play the stock market well.} i like to create websites."
    
    \textbf{Comments}: "Wow, this review blew my mind ! The way AI is changing how we use smartphones, especially for someone like me who's always on the lookout for new tech and easy ways to maximize efficiency, is fascinating. \textit{Tech like this could really shake up the stock market for tech companies!}"
\end{quote}

\section{Crowd-Sourced Comment Samples}

\label{app:SampledRealComments}
We demonstrate Persona-based comments, Real comments used for this evaluation and their scores.
Three score are listed in the order of \textbf{Relevance, Believability and Helpfulness}.
\subsection{Real Comments}
\begin{quote}
    \textbf{Video Title}: We Sent Garlic Bread to the Edge of Space, Then Ate It.
    
    \textbf{Comments}: That’s why aliens hate people. We act like we are giving them garlic bread and took it back. It makes them look like fool. 
    \textbf{Scores}: 4.32, 3.86, 3.5.
    
    \textbf{Comments}: Is it bird? Is it a plane? Nah... it’s a piece of garlic bread :D
    \textbf{Scores}: 5.36, 5.07, 4.18.
    
    \textbf{Comments}: STRATOSPHERIC BREAD! We just learned about this in science like last month ago.
    \textbf{Scores}: 4.79, 4.5, 4.18
\end{quote}

\subsection{Generated Comments}

\begin{quote}
    \textbf{Video Title}: We Sent Garlic Bread to the Edge of Space, Then Ate It

    \textbf{Comments}: Who else got here just to see if the space garlic bread tastes any different? 
    \textbf{Scores}: 5.39, 5.75, 4.75
    
    \textbf{Comments}: Nothing like combining the thrill of space with the comfort of garlic bread! This is the kind of adventurous spirit I live for. Great to see something so out-of-the-box! Makes me ponder what other foods could take a stratospheric journey. \#SpaceBreadAdventures 
    \textbf{Scores}: 5.64, 5.89, 5.14.

    \textbf{Comments}: Interesting experiment, but how about considering the environmental impact of such stunts? Launching objects into the stratosphere for taste tests seems a bit frivolous given the broader challenges our planet faces
    \textbf{Scores}: 4.68, 4.89, 5.54.

\end{quote}

%% file: main.bbl

\begin{thebibliography}{86}


\ifx \showCODEN    \undefined \def \showCODEN     #1{\unskip}     \fi
\ifx \showDOI      \undefined \def \showDOI       #1{#1}\fi
\ifx \showISBNx    \undefined \def \showISBNx     #1{\unskip}     \fi
\ifx \showISBNxiii \undefined \def \showISBNxiii  #1{\unskip}     \fi
\ifx \showISSN     \undefined \def \showISSN      #1{\unskip}     \fi
\ifx \showLCCN     \undefined \def \showLCCN      #1{\unskip}     \fi
\ifx \shownote     \undefined \def \shownote      #1{#1}          \fi
\ifx \showarticletitle \undefined \def \showarticletitle #1{#1}   \fi
\ifx \showURL      \undefined \def \showURL       {\relax}        \fi
\providecommand\bibfield[2]{#2}
\providecommand\bibinfo[2]{#2}
\providecommand\natexlab[1]{#1}
\providecommand\showeprint[2][]{arXiv:#2}

\bibitem[\protect\citeauthoryear{Antol, Agrawal, Lu, Mitchell, Batra, Zitnick, and Parikh}{Antol et~al\mbox{.}}{2015}]%
        {antol2015vqa}
\bibfield{author}{\bibinfo{person}{Stanislaw Antol}, \bibinfo{person}{Aishwarya Agrawal}, \bibinfo{person}{Jiasen Lu}, \bibinfo{person}{Margaret Mitchell}, \bibinfo{person}{Dhruv Batra}, \bibinfo{person}{C~Lawrence Zitnick}, {and} \bibinfo{person}{Devi Parikh}.} \bibinfo{year}{2015}\natexlab{}.
\newblock \showarticletitle{Vqa: Visual question answering}. In \bibinfo{booktitle}{\emph{Proceedings of the IEEE international conference on computer vision}}. \bibinfo{pages}{2425--2433}.
\newblock


\bibitem[\protect\citeauthoryear{{Arjun Panickssery and Samuel R. Bowman and Shi Feng}}{{Arjun Panickssery and Samuel R. Bowman and Shi Feng}}{2024}]%
        {panickssery2024llmevaluatorsrecognizefavor}
\bibfield{author}{\bibinfo{person}{{Arjun Panickssery and Samuel R. Bowman and Shi Feng}}.} \bibinfo{year}{2024}\natexlab{}.
\newblock \bibinfo{title}{LLM Evaluators Recognize and Favor Their Own Generations}.
\newblock
\newblock
\showeprint[arxiv]{2404.13076}~[cs.CL]
\urldef\tempurl%
\url{https://arxiv.org/abs/2404.13076}
\showURL{%
\tempurl}


\bibitem[\protect\citeauthoryear{Barbu, Bridge, Burchill, Coroian, Dickinson, Fidler, Michaux, Mussman, Narayanaswamy, Salvi, Schmidt, Shangguan, Siskind, Waggoner, Wang, Wei, Yin, and Zhang}{Barbu et~al\mbox{.}}{2012}]%
        {barbu2012videosentences}
\bibfield{author}{\bibinfo{person}{Andrei Barbu}, \bibinfo{person}{Alexander Bridge}, \bibinfo{person}{Zachary Burchill}, \bibinfo{person}{Dan Coroian}, \bibinfo{person}{Sven Dickinson}, \bibinfo{person}{Sanja Fidler}, \bibinfo{person}{Aaron Michaux}, \bibinfo{person}{Sam Mussman}, \bibinfo{person}{Siddharth Narayanaswamy}, \bibinfo{person}{Dhaval Salvi}, \bibinfo{person}{Lara Schmidt}, \bibinfo{person}{Jiangnan Shangguan}, \bibinfo{person}{Jeffrey~Mark Siskind}, \bibinfo{person}{Jarrell Waggoner}, \bibinfo{person}{Song Wang}, \bibinfo{person}{Jinlian Wei}, \bibinfo{person}{Yifan Yin}, {and} \bibinfo{person}{Zhiqi Zhang}.} \bibinfo{year}{2012}\natexlab{}.
\newblock \bibinfo{title}{Video In Sentences Out}.
\newblock
\newblock
\showeprint[arxiv]{1204.2742}~[cs.CV]
\urldef\tempurl%
\url{https://arxiv.org/abs/1204.2742}
\showURL{%
\tempurl}


\bibitem[\protect\citeauthoryear{Barnard}{Barnard}{2016}]%
        {Barnard2016VisionLanguage}
\bibfield{author}{\bibinfo{person}{Kobus Barnard}.} \bibinfo{year}{2016}\natexlab{}.
\newblock \showarticletitle{Computational Methods for Integrating Vision and Language}.
\newblock \bibinfo{journal}{\emph{Synthesis Lectures on Computer Vision}}  \bibinfo{volume}{6} (\bibinfo{date}{04} \bibinfo{year}{2016}), \bibinfo{pages}{1--227}.
\newblock
\urldef\tempurl%
\url{https://doi.org/10.2200/S00705ED1V01Y201602COV007}
\showDOI{\tempurl}


\bibitem[\protect\citeauthoryear{Benharrak, Zindulka, Lehmann, Heuer, and Buschek}{Benharrak et~al\mbox{.}}{2024}]%
        {Benharrak_2024}
\bibfield{author}{\bibinfo{person}{Karim Benharrak}, \bibinfo{person}{Tim Zindulka}, \bibinfo{person}{Florian Lehmann}, \bibinfo{person}{Hendrik Heuer}, {and} \bibinfo{person}{Daniel Buschek}.} \bibinfo{year}{2024}\natexlab{}.
\newblock \showarticletitle{Writer-Defined AI Personas for On-Demand Feedback Generation}. In \bibinfo{booktitle}{\emph{Proceedings of the CHI Conference on Human Factors in Computing Systems}} \emph{(\bibinfo{series}{CHI ’24}, Vol.~\bibinfo{volume}{11})}. \bibinfo{publisher}{ACM}, \bibinfo{pages}{1–18}.
\newblock
\urldef\tempurl%
\url{https://doi.org/10.1145/3613904.3642406}
\showDOI{\tempurl}


\bibitem[\protect\citeauthoryear{Bogolin, Croitoru, and Leordeanu}{Bogolin et~al\mbox{.}}{2020}]%
        {Bogolin2020VNLG}
\bibfield{author}{\bibinfo{person}{Simion-Vlad Bogolin}, \bibinfo{person}{Ioana Croitoru}, {and} \bibinfo{person}{Marius Leordeanu}.} \bibinfo{year}{2020}\natexlab{}.
\newblock \showarticletitle{A hierarchical approach to vision-based language generation: from simple sentences to complex natural language}. In \bibinfo{booktitle}{\emph{Proceedings of the 28th International Conference on Computational Linguistics}}, \bibfield{editor}{\bibinfo{person}{Donia Scott}, \bibinfo{person}{Nuria Bel}, {and} \bibinfo{person}{Chengqing Zong}} (Eds.). \bibinfo{publisher}{International Committee on Computational Linguistics}, \bibinfo{address}{Barcelona, Spain (Online)}, \bibinfo{pages}{2436--2447}.
\newblock
\urldef\tempurl%
\url{https://doi.org/10.18653/v1/2020.coling-main.220}
\showDOI{\tempurl}


\bibitem[\protect\citeauthoryear{Brown}{Brown}{2020}]%
        {brown2020language}
\bibfield{author}{\bibinfo{person}{Tom~B Brown}.} \bibinfo{year}{2020}\natexlab{}.
\newblock \showarticletitle{Language models are few-shot learners}.
\newblock \bibinfo{journal}{\emph{arXiv preprint arXiv:2005.14165}} (\bibinfo{year}{2020}).
\newblock


\bibitem[\protect\citeauthoryear{Bubeck, Chandrasekaran, Eldan, Gehrke, Horvitz, Kamar, Lee, Lee, Li, Lundberg, Nori, Palangi, Ribeiro, and Zhang}{Bubeck et~al\mbox{.}}{2023}]%
        {bubeck2023sparksartificialgeneralintelligence}
\bibfield{author}{\bibinfo{person}{Sébastien Bubeck}, \bibinfo{person}{Varun Chandrasekaran}, \bibinfo{person}{Ronen Eldan}, \bibinfo{person}{Johannes Gehrke}, \bibinfo{person}{Eric Horvitz}, \bibinfo{person}{Ece Kamar}, \bibinfo{person}{Peter Lee}, \bibinfo{person}{Yin~Tat Lee}, \bibinfo{person}{Yuanzhi Li}, \bibinfo{person}{Scott Lundberg}, \bibinfo{person}{Harsha Nori}, \bibinfo{person}{Hamid Palangi}, \bibinfo{person}{Marco~Tulio Ribeiro}, {and} \bibinfo{person}{Yi Zhang}.} \bibinfo{year}{2023}\natexlab{}.
\newblock \bibinfo{title}{Sparks of Artificial General Intelligence: Early experiments with GPT-4}.
\newblock
\newblock
\showeprint[arxiv]{2303.12712}~[cs.CL]
\urldef\tempurl%
\url{https://arxiv.org/abs/2303.12712}
\showURL{%
\tempurl}


\bibitem[\protect\citeauthoryear{Chandu, Prabhumoye, Salakhutdinov, and Black}{Chandu et~al\mbox{.}}{2019}]%
        {chandu-etal-2019-way}
\bibfield{author}{\bibinfo{person}{Khyathi Chandu}, \bibinfo{person}{Shrimai Prabhumoye}, \bibinfo{person}{Ruslan Salakhutdinov}, {and} \bibinfo{person}{Alan~W Black}.} \bibinfo{year}{2019}\natexlab{}.
\newblock \showarticletitle{{``}My Way of Telling a Story{''}: Persona based Grounded Story Generation}. In \bibinfo{booktitle}{\emph{Proceedings of the Second Workshop on Storytelling}}, \bibfield{editor}{\bibinfo{person}{Francis Ferraro}, \bibinfo{person}{Ting-Hao~{`}Kenneth{'} Huang}, \bibinfo{person}{Stephanie~M. Lukin}, {and} \bibinfo{person}{Margaret Mitchell}} (Eds.). \bibinfo{publisher}{Association for Computational Linguistics}, \bibinfo{address}{Florence, Italy}, \bibinfo{pages}{11--21}.
\newblock
\urldef\tempurl%
\url{https://doi.org/10.18653/v1/W19-3402}
\showDOI{\tempurl}


\bibitem[\protect\citeauthoryear{Chelaru, Orellana-Rodriguez, and Altingovde}{Chelaru et~al\mbox{.}}{2014}]%
        {Chelaru2014socialfeedback}
\bibfield{author}{\bibinfo{person}{Sorina Chelaru}, \bibinfo{person}{Claudia Orellana-Rodriguez}, {and} \bibinfo{person}{Ismail~Sengor Altingovde}.} \bibinfo{year}{2014}\natexlab{}.
\newblock \showarticletitle{How useful is social feedback for learning to rank YouTube videos?}
\newblock \bibinfo{journal}{\emph{World Wide Web}} \bibinfo{volume}{17}, \bibinfo{number}{5} (\bibinfo{year}{2014}), \bibinfo{pages}{997--1025}.
\newblock
\urldef\tempurl%
\url{https://doi.org/10.1007/s11280-013-0258-9}
\showDOI{\tempurl}


\bibitem[\protect\citeauthoryear{Chen, Wei, Li, Dong, Zhang, Zang, Chen, Duan, Lin, Tang, Yuan, Qiao, Lin, Zhao, and Wang}{Chen et~al\mbox{.}}{2024}]%
        {chen2024sharegpt4videoimprovingvideounderstanding}
\bibfield{author}{\bibinfo{person}{Lin Chen}, \bibinfo{person}{Xilin Wei}, \bibinfo{person}{Jinsong Li}, \bibinfo{person}{Xiaoyi Dong}, \bibinfo{person}{Pan Zhang}, \bibinfo{person}{Yuhang Zang}, \bibinfo{person}{Zehui Chen}, \bibinfo{person}{Haodong Duan}, \bibinfo{person}{Bin Lin}, \bibinfo{person}{Zhenyu Tang}, \bibinfo{person}{Li Yuan}, \bibinfo{person}{Yu Qiao}, \bibinfo{person}{Dahua Lin}, \bibinfo{person}{Feng Zhao}, {and} \bibinfo{person}{Jiaqi Wang}.} \bibinfo{year}{2024}\natexlab{}.
\newblock \bibinfo{title}{ShareGPT4Video: Improving Video Understanding and Generation with Better Captions}.
\newblock
\newblock
\showeprint[arxiv]{2406.04325}~[cs.CV]
\urldef\tempurl%
\url{https://arxiv.org/abs/2406.04325}
\showURL{%
\tempurl}


\bibitem[\protect\citeauthoryear{Colin~Dodds}{Colin~Dodds}{2024}]%
        {Colin2024Showandtell}
\bibfield{author}{\bibinfo{person}{Ahmed~Kharrufa Colin~Dodds}.} \bibinfo{year}{2024}\natexlab{}.
\newblock \showarticletitle{Show-and-Tell: An Interface for Delivering Rich Feedback upon Creative Media Artefacts}.
\newblock \bibinfo{journal}{\emph{Multimodal Technol. Interact}} (\bibinfo{year}{2024}).
\newblock
\urldef\tempurl%
\url{https://www.mdpi.com/2414-4088/8/3/23}
\showURL{%
\tempurl}


\bibitem[\protect\citeauthoryear{{Digital Marketing Institute}}{{Digital Marketing Institute}}{2024}]%
        {DMI2024growYT}
\bibfield{author}{\bibinfo{person}{{Digital Marketing Institute}}.} \bibinfo{year}{2024}\natexlab{}.
\newblock \bibinfo{title}{14 Ways to Grow Your YouTube Channel}.
\newblock \bibinfo{howpublished}{\url{https://digitalmarketinginstitute.com/}}.
\newblock
\urldef\tempurl%
\url{https://digitalmarketinginstitute.com/blog/10-ways-to-grow-your-youtube-channel-in-2018}
\showURL{%
\tempurl}
\newblock
\shownote{Accessed: 2024-10-01.}


\bibitem[\protect\citeauthoryear{Dognin, Melnyk, Mroueh, Padhi, Rigotti, Ross, Schiff, Young, and Belgodere}{Dognin et~al\mbox{.}}{2022}]%
        {dognin2022image}
\bibfield{author}{\bibinfo{person}{Pierre Dognin}, \bibinfo{person}{Igor Melnyk}, \bibinfo{person}{Youssef Mroueh}, \bibinfo{person}{Inkit Padhi}, \bibinfo{person}{Mattia Rigotti}, \bibinfo{person}{Jarret Ross}, \bibinfo{person}{Yair Schiff}, \bibinfo{person}{Richard~A Young}, {and} \bibinfo{person}{Brian Belgodere}.} \bibinfo{year}{2022}\natexlab{}.
\newblock \showarticletitle{Image captioning as an assistive technology: Lessons learned from vizwiz 2020 challenge}.
\newblock \bibinfo{journal}{\emph{Journal of Artificial Intelligence Research}}  \bibinfo{volume}{73} (\bibinfo{year}{2022}), \bibinfo{pages}{437--459}.
\newblock


\bibitem[\protect\citeauthoryear{Duan, Cui, Ma, Wei, Zhu, and Zhao}{Duan et~al\mbox{.}}{2020}]%
        {duan2020multimodalmatchingtransformerlive}
\bibfield{author}{\bibinfo{person}{Chaoqun Duan}, \bibinfo{person}{Lei Cui}, \bibinfo{person}{Shuming Ma}, \bibinfo{person}{Furu Wei}, \bibinfo{person}{Conghui Zhu}, {and} \bibinfo{person}{Tiejun Zhao}.} \bibinfo{year}{2020}\natexlab{}.
\newblock \bibinfo{title}{Multimodal Matching Transformer for Live Commenting}.
\newblock
\newblock
\showeprint[arxiv]{2002.02649}~[cs.CL]
\urldef\tempurl%
\url{https://arxiv.org/abs/2002.02649}
\showURL{%
\tempurl}


\bibitem[\protect\citeauthoryear{Dubovi and Tabak}{Dubovi and Tabak}{2020}]%
        {DUBOVI2020103939}
\bibfield{author}{\bibinfo{person}{Ilana Dubovi} {and} \bibinfo{person}{Iris Tabak}.} \bibinfo{year}{2020}\natexlab{}.
\newblock \showarticletitle{An empirical analysis of knowledge co-construction in YouTube comments}.
\newblock \bibinfo{journal}{\emph{Computers \& Education}}  \bibinfo{volume}{156} (\bibinfo{year}{2020}), \bibinfo{pages}{103939}.
\newblock
\showISSN{0360-1315}
\urldef\tempurl%
\url{https://doi.org/10.1016/j.compedu.2020.103939}
\showDOI{\tempurl}


\bibitem[\protect\citeauthoryear{Farhadi, Hejrati, Sadeghi, Young, Rashtchian, Hockenmaier, and Forsyth}{Farhadi et~al\mbox{.}}{2010}]%
        {Farhadi2010PictureStory}
\bibfield{author}{\bibinfo{person}{Ali Farhadi}, \bibinfo{person}{Mohsen Hejrati}, \bibinfo{person}{Mohammad~Amin Sadeghi}, \bibinfo{person}{Peter Young}, \bibinfo{person}{Cyrus Rashtchian}, \bibinfo{person}{Julia Hockenmaier}, {and} \bibinfo{person}{David Forsyth}.} \bibinfo{year}{2010}\natexlab{}.
\newblock \showarticletitle{Every Picture Tells a Story: Generating Sentences from Images}. In \bibinfo{booktitle}{\emph{Computer Vision -- ECCV 2010}}, \bibfield{editor}{\bibinfo{person}{Kostas Daniilidis}, \bibinfo{person}{Petros Maragos}, {and} \bibinfo{person}{Nikos Paragios}} (Eds.). \bibinfo{publisher}{Springer Berlin Heidelberg}, \bibinfo{address}{Berlin, Heidelberg}, \bibinfo{pages}{15--29}.
\newblock
\showISBNx{978-3-642-15561-1}


\bibitem[\protect\citeauthoryear{Gong, Li, Feng, Wu, and Kong}{Gong et~al\mbox{.}}{2022}]%
        {gong2022diffuseq}
\bibfield{author}{\bibinfo{person}{Shansan Gong}, \bibinfo{person}{Mukai Li}, \bibinfo{person}{Jiangtao Feng}, \bibinfo{person}{Zhiyong Wu}, {and} \bibinfo{person}{LingPeng Kong}.} \bibinfo{year}{2022}\natexlab{}.
\newblock \showarticletitle{Diffuseq: Sequence to sequence text generation with diffusion models}.
\newblock \bibinfo{journal}{\emph{arXiv preprint arXiv:2210.08933}} (\bibinfo{year}{2022}).
\newblock


\bibitem[\protect\citeauthoryear{Goyal, Li, and Durrett}{Goyal et~al\mbox{.}}{2022}]%
        {Goyal2022NewsSA}
\bibfield{author}{\bibinfo{person}{Tanya Goyal}, \bibinfo{person}{Junyi~Jessy Li}, {and} \bibinfo{person}{Greg Durrett}.} \bibinfo{year}{2022}\natexlab{}.
\newblock \showarticletitle{News Summarization and Evaluation in the Era of GPT-3}.
\newblock \bibinfo{journal}{\emph{ArXiv}}  \bibinfo{volume}{abs/2209.12356} (\bibinfo{year}{2022}).
\newblock
\urldef\tempurl%
\url{https://api.semanticscholar.org/CorpusID:252532176}
\showURL{%
\tempurl}


\bibitem[\protect\citeauthoryear{Goyal, Khot, Summers-Stay, Batra, and Parikh}{Goyal et~al\mbox{.}}{2017}]%
        {goyal2017making}
\bibfield{author}{\bibinfo{person}{Yash Goyal}, \bibinfo{person}{Tejas Khot}, \bibinfo{person}{Douglas Summers-Stay}, \bibinfo{person}{Dhruv Batra}, {and} \bibinfo{person}{Devi Parikh}.} \bibinfo{year}{2017}\natexlab{}.
\newblock \showarticletitle{Making the v in vqa matter: Elevating the role of image understanding in visual question answering}. In \bibinfo{booktitle}{\emph{Proceedings of the IEEE conference on computer vision and pattern recognition}}. \bibinfo{pages}{6904--6913}.
\newblock


\bibitem[\protect\citeauthoryear{Guadarrama, Krishnamoorthy, Malkarnenkar, Venugopalan, Mooney, Darrell, and Saenko}{Guadarrama et~al\mbox{.}}{2013}]%
        {Guadarrama2013Youtube2Text}
\bibfield{author}{\bibinfo{person}{Sergio Guadarrama}, \bibinfo{person}{Niveda Krishnamoorthy}, \bibinfo{person}{Girish Malkarnenkar}, \bibinfo{person}{Subhashini Venugopalan}, \bibinfo{person}{Raymond Mooney}, \bibinfo{person}{Trevor Darrell}, {and} \bibinfo{person}{Kate Saenko}.} \bibinfo{year}{2013}\natexlab{}.
\newblock \showarticletitle{YouTube2Text: Recognizing and Describing Arbitrary Activities Using Semantic Hierarchies and Zero-Shot Recognition}. In \bibinfo{booktitle}{\emph{2013 IEEE International Conference on Computer Vision}}. \bibinfo{pages}{2712--2719}.
\newblock
\urldef\tempurl%
\url{https://doi.org/10.1109/ICCV.2013.337}
\showDOI{\tempurl}


\bibitem[\protect\citeauthoryear{Gurari, Zhao, Zhang, and Bhattacharya}{Gurari et~al\mbox{.}}{2020}]%
        {gurari2020captioning}
\bibfield{author}{\bibinfo{person}{Danna Gurari}, \bibinfo{person}{Yinan Zhao}, \bibinfo{person}{Meng Zhang}, {and} \bibinfo{person}{Nilavra Bhattacharya}.} \bibinfo{year}{2020}\natexlab{}.
\newblock \showarticletitle{Captioning images taken by people who are blind}. In \bibinfo{booktitle}{\emph{Computer Vision--ECCV 2020: 16th European Conference, Glasgow, UK, August 23--28, 2020, Proceedings, Part XVII 16}}. Springer, \bibinfo{pages}{417--434}.
\newblock


\bibitem[\protect\citeauthoryear{H\"{a}m\"{a}l\"{a}inen, Tavast, and Kunnari}{H\"{a}m\"{a}l\"{a}inen et~al\mbox{.}}{2023}]%
        {hamalainen2023LLMHCI}
\bibfield{author}{\bibinfo{person}{Perttu H\"{a}m\"{a}l\"{a}inen}, \bibinfo{person}{Mikke Tavast}, {and} \bibinfo{person}{Anton Kunnari}.} \bibinfo{year}{2023}\natexlab{}.
\newblock \showarticletitle{Evaluating Large Language Models in Generating Synthetic HCI Research Data: a Case Study}. In \bibinfo{booktitle}{\emph{Proceedings of the 2023 CHI Conference on Human Factors in Computing Systems}} (Hamburg, Germany) \emph{(\bibinfo{series}{CHI '23})}. \bibinfo{publisher}{Association for Computing Machinery}, \bibinfo{address}{New York, NY, USA}, Article \bibinfo{articleno}{433}, \bibinfo{numpages}{19}~pages.
\newblock
\showISBNx{9781450394215}
\urldef\tempurl%
\url{https://doi.org/10.1145/3544548.3580688}
\showDOI{\tempurl}


\bibitem[\protect\citeauthoryear{Hattie and Timperley}{Hattie and Timperley}{2007}]%
        {John2007feedbackpower}
\bibfield{author}{\bibinfo{person}{John Hattie} {and} \bibinfo{person}{Helen Timperley}.} \bibinfo{year}{2007}\natexlab{}.
\newblock \showarticletitle{The Power of Feedback}.
\newblock \bibinfo{journal}{\emph{Review of Educational Research}} \bibinfo{volume}{77}, \bibinfo{number}{1} (\bibinfo{year}{2007}), \bibinfo{pages}{81--112}.
\newblock
\urldef\tempurl%
\url{https://doi.org/10.3102/003465430298487}
\showDOI{\tempurl}
\showeprint{https://doi.org/10.3102/003465430298487}


\bibitem[\protect\citeauthoryear{Holmbom}{Holmbom}{2015}]%
        {Holmbom2015YoutuberStudy}
\bibfield{author}{\bibinfo{person}{Maria Holmbom}.} \bibinfo{year}{2015}\natexlab{}.
\newblock \emph{\bibinfo{title}{The YouTuber: A Qualitative Study of Popular Content Creators}}.
\newblock Dissertation. \bibinfo{school}{Umeå University}.
\newblock
\urldef\tempurl%
\url{https://urn.kb.se/resolve?urn=urn:nbn:se:umu:diva-105388}
\showURL{%
\tempurl}


\bibitem[\protect\citeauthoryear{Hossain, Sohel, Shiratuddin, and Laga}{Hossain et~al\mbox{.}}{2019}]%
        {hossain2019ImageCaption}
\bibfield{author}{\bibinfo{person}{MD.~Zakir Hossain}, \bibinfo{person}{Ferdous Sohel}, \bibinfo{person}{Mohd~Fairuz Shiratuddin}, {and} \bibinfo{person}{Hamid Laga}.} \bibinfo{year}{2019}\natexlab{}.
\newblock \showarticletitle{A Comprehensive Survey of Deep Learning for Image Captioning}.
\newblock \bibinfo{journal}{\emph{ACM Comput. Surv.}} \bibinfo{volume}{51}, \bibinfo{number}{6}, Article \bibinfo{articleno}{118} (\bibinfo{date}{feb} \bibinfo{year}{2019}), \bibinfo{numpages}{36}~pages.
\newblock
\showISSN{0360-0300}
\urldef\tempurl%
\url{https://doi.org/10.1145/3295748}
\showDOI{\tempurl}


\bibitem[\protect\citeauthoryear{Huang, Ramanathan, Mahajan, Torresani, Paluri, Fei-Fei, and Niebles}{Huang et~al\mbox{.}}{2018}]%
        {Huang_2018_CVPR}
\bibfield{author}{\bibinfo{person}{De-An Huang}, \bibinfo{person}{Vignesh Ramanathan}, \bibinfo{person}{Dhruv Mahajan}, \bibinfo{person}{Lorenzo Torresani}, \bibinfo{person}{Manohar Paluri}, \bibinfo{person}{Li Fei-Fei}, {and} \bibinfo{person}{Juan~Carlos Niebles}.} \bibinfo{year}{2018}\natexlab{}.
\newblock \showarticletitle{What Makes a Video a Video: Analyzing Temporal Information in Video Understanding Models and Datasets}. In \bibinfo{booktitle}{\emph{Proceedings of the IEEE Conference on Computer Vision and Pattern Recognition (CVPR)}}.
\newblock


\bibitem[\protect\citeauthoryear{Huang, Ferraro, Mostafazadeh, Misra, Agrawal, Devlin, Girshick, He, Kohli, Batra, Zitnick, Parikh, Vanderwende, Galley, and Mitchell}{Huang et~al\mbox{.}}{2016}]%
        {huang-etal-2016-visual}
\bibfield{author}{\bibinfo{person}{Ting-Hao~Kenneth Huang}, \bibinfo{person}{Francis Ferraro}, \bibinfo{person}{Nasrin Mostafazadeh}, \bibinfo{person}{Ishan Misra}, \bibinfo{person}{Aishwarya Agrawal}, \bibinfo{person}{Jacob Devlin}, \bibinfo{person}{Ross Girshick}, \bibinfo{person}{Xiaodong He}, \bibinfo{person}{Pushmeet Kohli}, \bibinfo{person}{Dhruv Batra}, \bibinfo{person}{C.~Lawrence Zitnick}, \bibinfo{person}{Devi Parikh}, \bibinfo{person}{Lucy Vanderwende}, \bibinfo{person}{Michel Galley}, {and} \bibinfo{person}{Margaret Mitchell}.} \bibinfo{year}{2016}\natexlab{}.
\newblock \showarticletitle{Visual Storytelling}. In \bibinfo{booktitle}{\emph{Proceedings of the 2016 Conference of the North {A}merican Chapter of the Association for Computational Linguistics: Human Language Technologies}}, \bibfield{editor}{\bibinfo{person}{Kevin Knight}, \bibinfo{person}{Ani Nenkova}, {and} \bibinfo{person}{Owen Rambow}} (Eds.). \bibinfo{publisher}{Association for Computational Linguistics}, \bibinfo{address}{San Diego, California}, \bibinfo{pages}{1233--1239}.
\newblock
\urldef\tempurl%
\url{https://doi.org/10.18653/v1/N16-1147}
\showDOI{\tempurl}


\bibitem[\protect\citeauthoryear{Huh, Yang, and Peng}{Huh et~al\mbox{.}}{[n.d.]}]%
        {huh2023xiang}
\bibfield{author}{\bibinfo{person}{Mina Huh}, \bibinfo{person}{Saelyne Yang}, {and} \bibinfo{person}{Yi-Hao Peng}.} \bibinfo{year}{[n.d.]}\natexlab{}.
\newblock \showarticletitle{Xiang’Anthony’Chen, Young-Ho Kim, and Amy Pavel. 2023. AVscript: Accessible Video Editing with Audio-Visual Scripts}. In \bibinfo{booktitle}{\emph{Proceedings of the 2023 CHI Conference on Human Factors in Computing Systems}}. \bibinfo{pages}{1--17}.
\newblock


\bibitem[\protect\citeauthoryear{Jiang, Zhang, Cao, Breazeal, Roy, and Kabbara}{Jiang et~al\mbox{.}}{2024}]%
        {jiang2024personallminvestigatingabilitylarge}
\bibfield{author}{\bibinfo{person}{Hang Jiang}, \bibinfo{person}{Xiajie Zhang}, \bibinfo{person}{Xubo Cao}, \bibinfo{person}{Cynthia Breazeal}, \bibinfo{person}{Deb Roy}, {and} \bibinfo{person}{Jad Kabbara}.} \bibinfo{year}{2024}\natexlab{}.
\newblock \bibinfo{title}{PersonaLLM: Investigating the Ability of Large Language Models to Express Personality Traits}.
\newblock
\newblock
\showeprint[arxiv]{2305.02547}~[cs.CL]
\urldef\tempurl%
\url{https://arxiv.org/abs/2305.02547}
\showURL{%
\tempurl}


\bibitem[\protect\citeauthoryear{josepmartins}{josepmartins}{[n.d.]}]%
        {2023BoringAvatar}
\bibfield{author}{\bibinfo{person}{josepmartins}.} \bibinfo{year}{[n.d.]}\natexlab{}.
\newblock \bibinfo{title}{boring-avatar}.
\newblock
\newblock
\urldef\tempurl%
\url{https://github.com/boringdesigners/boring-avatars}
\showURL{%
\tempurl}


\bibitem[\protect\citeauthoryear{Kim, Heo, Son, Park, and Zhang}{Kim et~al\mbox{.}}{2019}]%
        {kim2019glacnetglocalattention}
\bibfield{author}{\bibinfo{person}{Taehyeong Kim}, \bibinfo{person}{Min-Oh Heo}, \bibinfo{person}{Seonil Son}, \bibinfo{person}{Kyoung-Wha Park}, {and} \bibinfo{person}{Byoung-Tak Zhang}.} \bibinfo{year}{2019}\natexlab{}.
\newblock \bibinfo{title}{GLAC Net: GLocal Attention Cascading Networks for Multi-image Cued Story Generation}.
\newblock
\newblock
\showeprint[arxiv]{1805.10973}~[cs.CL]
\urldef\tempurl%
\url{https://arxiv.org/abs/1805.10973}
\showURL{%
\tempurl}


\bibitem[\protect\citeauthoryear{Kluger and DeNisi}{Kluger and DeNisi}{1996}]%
        {kluger1996effects}
\bibfield{author}{\bibinfo{person}{Avraham~N Kluger} {and} \bibinfo{person}{Angelo DeNisi}.} \bibinfo{year}{1996}\natexlab{}.
\newblock \showarticletitle{The effects of feedback interventions on performance: a historical review, a meta-analysis, and a preliminary feedback intervention theory.}
\newblock \bibinfo{journal}{\emph{Psychological bulletin}} \bibinfo{volume}{119}, \bibinfo{number}{2} (\bibinfo{year}{1996}), \bibinfo{pages}{254}.
\newblock


\bibitem[\protect\citeauthoryear{Kolasani}{Kolasani}{2023}]%
        {Saydulu2023Customer}
\bibfield{author}{\bibinfo{person}{Saydulu Kolasani}.} \bibinfo{year}{2023}\natexlab{}.
\newblock \showarticletitle{Optimizing Natural Language Processing, Large Language Models (LLMs) for Efficient Customer Service, and hyper-personalization to enable sustainable growth and revenue}.
\newblock \bibinfo{journal}{\emph{Transactions on Latest Trends in Artificial Intelligence}} \bibinfo{volume}{4}, \bibinfo{number}{4} (\bibinfo{year}{2023}).
\newblock
\showISSN{3246-548X}
\urldef\tempurl%
\url{https://ijsdcs.com/index.php/TLAI/article/view/476}
\showURL{%
\tempurl}


\bibitem[\protect\citeauthoryear{Kulkarni, Bernstein, and Klemmer}{Kulkarni et~al\mbox{.}}{2015}]%
        {chinmaye2015Peerstudio}
\bibfield{author}{\bibinfo{person}{Chinmay~E. Kulkarni}, \bibinfo{person}{Michael~S. Bernstein}, {and} \bibinfo{person}{Scott~R. Klemmer}.} \bibinfo{year}{2015}\natexlab{}.
\newblock \showarticletitle{PeerStudio: Rapid Peer Feedback Emphasizes Revision and Improves Performance}. In \bibinfo{booktitle}{\emph{Proceedings of the Second (2015) ACM Conference on Learning @ Scale}} (Vancouver, BC, Canada) \emph{(\bibinfo{series}{L@S '15})}. \bibinfo{publisher}{Association for Computing Machinery}, \bibinfo{address}{New York, NY, USA}, \bibinfo{pages}{75–84}.
\newblock
\showISBNx{9781450334112}
\urldef\tempurl%
\url{https://doi.org/10.1145/2724660.2724670}
\showDOI{\tempurl}


\bibitem[\protect\citeauthoryear{Kulkarni, Premraj, Dhar, Li, Choi, Berg, and Berg}{Kulkarni et~al\mbox{.}}{2011}]%
        {Kulkarni2011BabyTalk}
\bibfield{author}{\bibinfo{person}{Girish Kulkarni}, \bibinfo{person}{Visruth Premraj}, \bibinfo{person}{Sagnik Dhar}, \bibinfo{person}{Siming Li}, \bibinfo{person}{Yejin Choi}, \bibinfo{person}{Alexander~C Berg}, {and} \bibinfo{person}{Tamara~L Berg}.} \bibinfo{year}{2011}\natexlab{}.
\newblock \showarticletitle{Baby talk: Understanding and generating simple image descriptions}. In \bibinfo{booktitle}{\emph{CVPR 2011}}. \bibinfo{pages}{1601--1608}.
\newblock
\urldef\tempurl%
\url{https://doi.org/10.1109/CVPR.2011.5995466}
\showDOI{\tempurl}


\bibitem[\protect\citeauthoryear{Kuznetsova, Ordonez, Berg, Berg, and Choi}{Kuznetsova et~al\mbox{.}}{2012}]%
        {kuznetsova-etal-2012-collective}
\bibfield{author}{\bibinfo{person}{Polina Kuznetsova}, \bibinfo{person}{Vicente Ordonez}, \bibinfo{person}{Alexander Berg}, \bibinfo{person}{Tamara Berg}, {and} \bibinfo{person}{Yejin Choi}.} \bibinfo{year}{2012}\natexlab{}.
\newblock \showarticletitle{Collective Generation of Natural Image Descriptions}. In \bibinfo{booktitle}{\emph{Proceedings of the 50th Annual Meeting of the Association for Computational Linguistics (Volume 1: Long Papers)}}, \bibfield{editor}{\bibinfo{person}{Haizhou Li}, \bibinfo{person}{Chin-Yew Lin}, \bibinfo{person}{Miles Osborne}, \bibinfo{person}{Gary~Geunbae Lee}, {and} \bibinfo{person}{Jong~C. Park}} (Eds.). \bibinfo{publisher}{Association for Computational Linguistics}, \bibinfo{address}{Jeju Island, Korea}, \bibinfo{pages}{359--368}.
\newblock
\urldef\tempurl%
\url{https://aclanthology.org/P12-1038}
\showURL{%
\tempurl}


\bibitem[\protect\citeauthoryear{Lee, Liang, and Yang}{Lee et~al\mbox{.}}{2022}]%
        {coauthor2022}
\bibfield{author}{\bibinfo{person}{Mina Lee}, \bibinfo{person}{Percy Liang}, {and} \bibinfo{person}{Qian Yang}.} \bibinfo{year}{2022}\natexlab{}.
\newblock \showarticletitle{CoAuthor: Designing a Human-AI Collaborative Writing Dataset for Exploring Language Model Capabilities}. In \bibinfo{booktitle}{\emph{Proceedings of the 2022 CHI Conference on Human Factors in Computing Systems}} (New Orleans, LA, USA) \emph{(\bibinfo{series}{CHI '22})}. \bibinfo{publisher}{Association for Computing Machinery}, \bibinfo{address}{New York, NY, USA}, Article \bibinfo{articleno}{388}, \bibinfo{numpages}{19}~pages.
\newblock
\showISBNx{9781450391573}
\urldef\tempurl%
\url{https://doi.org/10.1145/3491102.3502030}
\showDOI{\tempurl}


\bibitem[\protect\citeauthoryear{Lee and Ko}{Lee and Ko}{2011}]%
        {Lee2011Programming}
\bibfield{author}{\bibinfo{person}{Michael~J. Lee} {and} \bibinfo{person}{Amy~J. Ko}.} \bibinfo{year}{2011}\natexlab{}.
\newblock \showarticletitle{Personifying programming tool feedback improves novice programmers' learning}. In \bibinfo{booktitle}{\emph{Proceedings of the Seventh International Workshop on Computing Education Research}} (Providence, Rhode Island, USA) \emph{(\bibinfo{series}{ICER '11})}. \bibinfo{publisher}{Association for Computing Machinery}, \bibinfo{address}{New York, NY, USA}, \bibinfo{pages}{109–116}.
\newblock
\showISBNx{9781450308298}
\urldef\tempurl%
\url{https://doi.org/10.1145/2016911.2016934}
\showDOI{\tempurl}


\bibitem[\protect\citeauthoryear{Li, Galley, Brockett, Gao, and Dolan}{Li et~al\mbox{.}}{2016}]%
        {li-etal-2016-diversity}
\bibfield{author}{\bibinfo{person}{Jiwei Li}, \bibinfo{person}{Michel Galley}, \bibinfo{person}{Chris Brockett}, \bibinfo{person}{Jianfeng Gao}, {and} \bibinfo{person}{Bill Dolan}.} \bibinfo{year}{2016}\natexlab{}.
\newblock \showarticletitle{A Diversity-Promoting Objective Function for Neural Conversation Models}. In \bibinfo{booktitle}{\emph{Proceedings of the 2016 Conference of the North {A}merican Chapter of the Association for Computational Linguistics: Human Language Technologies}}, \bibfield{editor}{\bibinfo{person}{Kevin Knight}, \bibinfo{person}{Ani Nenkova}, {and} \bibinfo{person}{Owen Rambow}} (Eds.). \bibinfo{publisher}{Association for Computational Linguistics}, \bibinfo{address}{San Diego, California}, \bibinfo{pages}{110--119}.
\newblock
\urldef\tempurl%
\url{https://doi.org/10.18653/v1/N16-1014}
\showDOI{\tempurl}


\bibitem[\protect\citeauthoryear{Lin}{Lin}{2004}]%
        {lin2004rouge}
\bibfield{author}{\bibinfo{person}{Chin-Yew Lin}.} \bibinfo{year}{2004}\natexlab{}.
\newblock \showarticletitle{Rouge: A package for automatic evaluation of summaries}. In \bibinfo{booktitle}{\emph{Text summarization branches out}}. \bibinfo{pages}{74--81}.
\newblock


\bibitem[\protect\citeauthoryear{Lin and Chen}{Lin and Chen}{2023}]%
        {lin-chen-2023-llm}
\bibfield{author}{\bibinfo{person}{Yen-Ting Lin} {and} \bibinfo{person}{Yun-Nung Chen}.} \bibinfo{year}{2023}\natexlab{}.
\newblock \showarticletitle{{LLM}-Eval: Unified Multi-Dimensional Automatic Evaluation for Open-Domain Conversations with Large Language Models}. In \bibinfo{booktitle}{\emph{Proceedings of the 5th Workshop on NLP for Conversational AI (NLP4ConvAI 2023)}}, \bibfield{editor}{\bibinfo{person}{Yun-Nung Chen} {and} \bibinfo{person}{Abhinav Rastogi}} (Eds.). \bibinfo{publisher}{Association for Computational Linguistics}, \bibinfo{address}{Toronto, Canada}, \bibinfo{pages}{47--58}.
\newblock
\urldef\tempurl%
\url{https://doi.org/10.18653/v1/2023.nlp4convai-1.5}
\showDOI{\tempurl}


\bibitem[\protect\citeauthoryear{Liu, Iter, Xu, Wang, Xu, and Zhu}{Liu et~al\mbox{.}}{2023}]%
        {liu2023g}
\bibfield{author}{\bibinfo{person}{Yang Liu}, \bibinfo{person}{Dan Iter}, \bibinfo{person}{Yichong Xu}, \bibinfo{person}{Shuohang Wang}, \bibinfo{person}{Ruochen Xu}, {and} \bibinfo{person}{Chenguang Zhu}.} \bibinfo{year}{2023}\natexlab{}.
\newblock \showarticletitle{G-eval: Nlg evaluation using gpt-4 with better human alignment}.
\newblock \bibinfo{journal}{\emph{arXiv preprint arXiv:2303.16634}} (\bibinfo{year}{2023}).
\newblock


\bibitem[\protect\citeauthoryear{Lu, Mishra, Xia, Qiu, Chang, Zhu, Tafjord, Clark, and Kalyan}{Lu et~al\mbox{.}}{2022}]%
        {lu2022learn}
\bibfield{author}{\bibinfo{person}{Pan Lu}, \bibinfo{person}{Swaroop Mishra}, \bibinfo{person}{Tanglin Xia}, \bibinfo{person}{Liang Qiu}, \bibinfo{person}{Kai-Wei Chang}, \bibinfo{person}{Song-Chun Zhu}, \bibinfo{person}{Oyvind Tafjord}, \bibinfo{person}{Peter Clark}, {and} \bibinfo{person}{Ashwin Kalyan}.} \bibinfo{year}{2022}\natexlab{}.
\newblock \showarticletitle{Learn to explain: Multimodal reasoning via thought chains for science question answering}.
\newblock \bibinfo{journal}{\emph{Advances in Neural Information Processing Systems}}  \bibinfo{volume}{35} (\bibinfo{year}{2022}), \bibinfo{pages}{2507--2521}.
\newblock


\bibitem[\protect\citeauthoryear{Ma, Cui, Dai, Wei, and Sun}{Ma et~al\mbox{.}}{2018}]%
        {ma2018livebotgeneratinglivevideo}
\bibfield{author}{\bibinfo{person}{Shuming Ma}, \bibinfo{person}{Lei Cui}, \bibinfo{person}{Damai Dai}, \bibinfo{person}{Furu Wei}, {and} \bibinfo{person}{Xu Sun}.} \bibinfo{year}{2018}\natexlab{}.
\newblock \bibinfo{title}{LiveBot: Generating Live Video Comments Based on Visual and Textual Contexts}.
\newblock
\newblock
\showeprint[arxiv]{1809.04938}~[cs.CL]
\urldef\tempurl%
\url{https://arxiv.org/abs/1809.04938}
\showURL{%
\tempurl}


\bibitem[\protect\citeauthoryear{Malik, Wu, Vasavada, Song, Coots, Mitchell, Goodman, and Piech}{Malik et~al\mbox{.}}{2021}]%
        {malik2021generativegradingnearhumanlevel}
\bibfield{author}{\bibinfo{person}{Ali Malik}, \bibinfo{person}{Mike Wu}, \bibinfo{person}{Vrinda Vasavada}, \bibinfo{person}{Jinpeng Song}, \bibinfo{person}{Madison Coots}, \bibinfo{person}{John Mitchell}, \bibinfo{person}{Noah Goodman}, {and} \bibinfo{person}{Chris Piech}.} \bibinfo{year}{2021}\natexlab{}.
\newblock \bibinfo{title}{Generative Grading: Near Human-level Accuracy for Automated Feedback on Richly Structured Problems}.
\newblock
\newblock
\showeprint[arxiv]{1905.09916}~[cs.LG]
\urldef\tempurl%
\url{https://arxiv.org/abs/1905.09916}
\showURL{%
\tempurl}


\bibitem[\protect\citeauthoryear{{Meta}}{{Meta}}{2022}]%
        {personachat2022}
\bibfield{author}{\bibinfo{person}{{Meta}}.} \bibinfo{year}{2022}\natexlab{}.
\newblock \bibinfo{title}{PersonaChat}.
\newblock
\newblock
\urldef\tempurl%
\url{https://www.kaggle.com/datasets/atharvjairath/personachat}
\showURL{%
\tempurl}
\newblock
\shownote{Accessed: 2024-10-01.}


\bibitem[\protect\citeauthoryear{Mirowski, Mathewson, Pittman, and Evans}{Mirowski et~al\mbox{.}}{2023}]%
        {dramatron2023}
\bibfield{author}{\bibinfo{person}{Piotr Mirowski}, \bibinfo{person}{Kory~W. Mathewson}, \bibinfo{person}{Jaylen Pittman}, {and} \bibinfo{person}{Richard Evans}.} \bibinfo{year}{2023}\natexlab{}.
\newblock \showarticletitle{Co-Writing Screenplays and Theatre Scripts with Language Models: Evaluation by Industry Professionals}. In \bibinfo{booktitle}{\emph{Proceedings of the 2023 CHI Conference on Human Factors in Computing Systems}} (Hamburg, Germany) \emph{(\bibinfo{series}{CHI '23})}. \bibinfo{publisher}{Association for Computing Machinery}, \bibinfo{address}{New York, NY, USA}, Article \bibinfo{articleno}{355}, \bibinfo{numpages}{34}~pages.
\newblock
\showISBNx{9781450394215}
\urldef\tempurl%
\url{https://doi.org/10.1145/3544548.3581225}
\showDOI{\tempurl}


\bibitem[\protect\citeauthoryear{Mitchell, Han, Dodge, Mensch, Goyal, Berg, Yamaguchi, Berg, Stratos, and Daum\'{e}}{Mitchell et~al\mbox{.}}{2012}]%
        {mitchell2012midge}
\bibfield{author}{\bibinfo{person}{Margaret Mitchell}, \bibinfo{person}{Xufeng Han}, \bibinfo{person}{Jesse Dodge}, \bibinfo{person}{Alyssa Mensch}, \bibinfo{person}{Amit Goyal}, \bibinfo{person}{Alex Berg}, \bibinfo{person}{Kota Yamaguchi}, \bibinfo{person}{Tamara Berg}, \bibinfo{person}{Karl Stratos}, {and} \bibinfo{person}{Hal Daum\'{e}}.} \bibinfo{year}{2012}\natexlab{}.
\newblock \showarticletitle{Midge: generating image descriptions from computer vision detections}. In \bibinfo{booktitle}{\emph{Proceedings of the 13th Conference of the European Chapter of the Association for Computational Linguistics}} (Avignon, France) \emph{(\bibinfo{series}{EACL '12})}. \bibinfo{publisher}{Association for Computational Linguistics}, \bibinfo{address}{USA}, \bibinfo{pages}{747–756}.
\newblock
\showISBNx{9781937284190}


\bibitem[\protect\citeauthoryear{Nilforoshan and Wu}{Nilforoshan and Wu}{2018}]%
        {Nilforoshan_Wu_2018}
\bibfield{author}{\bibinfo{person}{Hamed Nilforoshan} {and} \bibinfo{person}{Eugene Wu}.} \bibinfo{year}{2018}\natexlab{}.
\newblock \showarticletitle{Leveraging Quality Prediction Models for Automatic Writing Feedback}.
\newblock \bibinfo{journal}{\emph{Proceedings of the International AAAI Conference on Web and Social Media}} \bibinfo{volume}{12}, \bibinfo{number}{1} (\bibinfo{date}{Jun.} \bibinfo{year}{2018}).
\newblock
\urldef\tempurl%
\url{https://doi.org/10.1609/icwsm.v12i1.14998}
\showDOI{\tempurl}


\bibitem[\protect\citeauthoryear{OpenAI, Achiam, Adler, Agarwal, Ahmad, Akkaya, Aleman, Almeida, Altenschmidt, Altman, Anadkat, Avila, Babuschkin, Balaji, Balcom, Baltescu, Bao, Bavarian, Belgum, Bello, Berdine, Bernadett-Shapiro, Berner, Bogdonoff, Boiko, Boyd, Brakman, Brockman, Brooks, Brundage, Button, Cai, Campbell, Cann, Carey, Carlson, Carmichael, Chan, Chang, Chantzis, Chen, Chen, Chen, Chen, Chen, Chess, Cho, Chu, Chung, Cummings, Currier, Dai, Decareaux, Degry, Deutsch, Deville, Dhar, Dohan, Dowling, Dunning, Ecoffet, Eleti, Eloundou, Farhi, Fedus, Felix, Fishman, Forte, Fulford, Gao, Georges, Gibson, Goel, Gogineni, Goh, Gontijo-Lopes, Gordon, Grafstein, Gray, Greene, Gross, Gu, Guo, Hallacy, Han, Harris, He, Heaton, Heidecke, Hesse, Hickey, Hickey, Hoeschele, Houghton, Hsu, Hu, Hu, Huizinga, Jain, Jain, Jang, Jiang, Jiang, Jin, Jin, Jomoto, Jonn, Jun, Kaftan, Łukasz Kaiser, Kamali, Kanitscheider, Keskar, Khan, Kilpatrick, Kim, Kim, Kim, Kirchner, Kiros, Knight, Kokotajlo, Łukasz Kondraciuk,
  Kondrich, Konstantinidis, Kosic, Krueger, Kuo, Lampe, Lan, Lee, Leike, Leung, Levy, Li, Lim, Lin, Lin, Litwin, Lopez, Lowe, Lue, Makanju, Malfacini, Manning, Markov, Markovski, Martin, Mayer, Mayne, McGrew, McKinney, McLeavey, McMillan, McNeil, Medina, Mehta, Menick, Metz, Mishchenko, Mishkin, Monaco, Morikawa, Mossing, Mu, Murati, Murk, Mély, Nair, Nakano, Nayak, Neelakantan, Ngo, Noh, Ouyang, O'Keefe, Pachocki, Paino, Palermo, Pantuliano, Parascandolo, Parish, Parparita, Passos, Pavlov, Peng, Perelman, de~Avila Belbute~Peres, Petrov, de~Oliveira~Pinto, Michael, Pokorny, Pokrass, Pong, Powell, Power, Power, Proehl, Puri, Radford, Rae, Ramesh, Raymond, Real, Rimbach, Ross, Rotsted, Roussez, Ryder, Saltarelli, Sanders, Santurkar, Sastry, Schmidt, Schnurr, Schulman, Selsam, Sheppard, Sherbakov, Shieh, Shoker, Shyam, Sidor, Sigler, Simens, Sitkin, Slama, Sohl, Sokolowsky, Song, Staudacher, Such, Summers, Sutskever, Tang, Tezak, Thompson, Tillet, Tootoonchian, Tseng, Tuggle, Turley, Tworek, Uribe, Vallone,
  Vijayvergiya, Voss, Wainwright, Wang, Wang, Wang, Ward, Wei, Weinmann, Welihinda, Welinder, Weng, Weng, Wiethoff, Willner, Winter, Wolrich, Wong, Workman, Wu, Wu, Wu, Xiao, Xu, Yoo, Yu, Yuan, Zaremba, Zellers, Zhang, Zhang, Zhao, Zheng, Zhuang, Zhuk, and Zoph}{OpenAI et~al\mbox{.}}{2023}]%
        {openai2023gpt4}
\bibfield{author}{\bibinfo{person}{OpenAI}, \bibinfo{person}{Josh Achiam}, \bibinfo{person}{Steven Adler}, \bibinfo{person}{Sandhini Agarwal}, \bibinfo{person}{Lama Ahmad}, \bibinfo{person}{Ilge Akkaya}, \bibinfo{person}{Florencia~Leoni Aleman}, \bibinfo{person}{Diogo Almeida}, \bibinfo{person}{Janko Altenschmidt}, \bibinfo{person}{Sam Altman}, \bibinfo{person}{Shyamal Anadkat}, \bibinfo{person}{Red Avila}, \bibinfo{person}{Igor Babuschkin}, \bibinfo{person}{Suchir Balaji}, \bibinfo{person}{Valerie Balcom}, \bibinfo{person}{Paul Baltescu}, \bibinfo{person}{Haiming Bao}, \bibinfo{person}{Mohammad Bavarian}, \bibinfo{person}{Jeff Belgum}, \bibinfo{person}{Irwan Bello}, \bibinfo{person}{Jake Berdine}, \bibinfo{person}{Gabriel Bernadett-Shapiro}, \bibinfo{person}{Christopher Berner}, \bibinfo{person}{Lenny Bogdonoff}, \bibinfo{person}{Oleg Boiko}, \bibinfo{person}{Madelaine Boyd}, \bibinfo{person}{Anna-Luisa Brakman}, \bibinfo{person}{Greg Brockman}, \bibinfo{person}{Tim Brooks}, \bibinfo{person}{Miles Brundage},
  \bibinfo{person}{Kevin Button}, \bibinfo{person}{Trevor Cai}, \bibinfo{person}{Rosie Campbell}, \bibinfo{person}{Andrew Cann}, \bibinfo{person}{Brittany Carey}, \bibinfo{person}{Chelsea Carlson}, \bibinfo{person}{Rory Carmichael}, \bibinfo{person}{Brooke Chan}, \bibinfo{person}{Che Chang}, \bibinfo{person}{Fotis Chantzis}, \bibinfo{person}{Derek Chen}, \bibinfo{person}{Sully Chen}, \bibinfo{person}{Ruby Chen}, \bibinfo{person}{Jason Chen}, \bibinfo{person}{Mark Chen}, \bibinfo{person}{Ben Chess}, \bibinfo{person}{Chester Cho}, \bibinfo{person}{Casey Chu}, \bibinfo{person}{Hyung~Won Chung}, \bibinfo{person}{Dave Cummings}, \bibinfo{person}{Jeremiah Currier}, \bibinfo{person}{Yunxing Dai}, \bibinfo{person}{Cory Decareaux}, \bibinfo{person}{Thomas Degry}, \bibinfo{person}{Noah Deutsch}, \bibinfo{person}{Damien Deville}, \bibinfo{person}{Arka Dhar}, \bibinfo{person}{David Dohan}, \bibinfo{person}{Steve Dowling}, \bibinfo{person}{Sheila Dunning}, \bibinfo{person}{Adrien Ecoffet}, \bibinfo{person}{Atty Eleti},
  \bibinfo{person}{Tyna Eloundou}, \bibinfo{person}{David Farhi}, \bibinfo{person}{Liam Fedus}, \bibinfo{person}{Niko Felix}, \bibinfo{person}{Simón~Posada Fishman}, \bibinfo{person}{Juston Forte}, \bibinfo{person}{Isabella Fulford}, \bibinfo{person}{Leo Gao}, \bibinfo{person}{Elie Georges}, \bibinfo{person}{Christian Gibson}, \bibinfo{person}{Vik Goel}, \bibinfo{person}{Tarun Gogineni}, \bibinfo{person}{Gabriel Goh}, \bibinfo{person}{Rapha Gontijo-Lopes}, \bibinfo{person}{Jonathan Gordon}, \bibinfo{person}{Morgan Grafstein}, \bibinfo{person}{Scott Gray}, \bibinfo{person}{Ryan Greene}, \bibinfo{person}{Joshua Gross}, \bibinfo{person}{Shixiang~Shane Gu}, \bibinfo{person}{Yufei Guo}, \bibinfo{person}{Chris Hallacy}, \bibinfo{person}{Jesse Han}, \bibinfo{person}{Jeff Harris}, \bibinfo{person}{Yuchen He}, \bibinfo{person}{Mike Heaton}, \bibinfo{person}{Johannes Heidecke}, \bibinfo{person}{Chris Hesse}, \bibinfo{person}{Alan Hickey}, \bibinfo{person}{Wade Hickey}, \bibinfo{person}{Peter Hoeschele},
  \bibinfo{person}{Brandon Houghton}, \bibinfo{person}{Kenny Hsu}, \bibinfo{person}{Shengli Hu}, \bibinfo{person}{Xin Hu}, \bibinfo{person}{Joost Huizinga}, \bibinfo{person}{Shantanu Jain}, \bibinfo{person}{Shawn Jain}, \bibinfo{person}{Joanne Jang}, \bibinfo{person}{Angela Jiang}, \bibinfo{person}{Roger Jiang}, \bibinfo{person}{Haozhun Jin}, \bibinfo{person}{Denny Jin}, \bibinfo{person}{Shino Jomoto}, \bibinfo{person}{Billie Jonn}, \bibinfo{person}{Heewoo Jun}, \bibinfo{person}{Tomer Kaftan}, \bibinfo{person}{Łukasz Kaiser}, \bibinfo{person}{Ali Kamali}, \bibinfo{person}{Ingmar Kanitscheider}, \bibinfo{person}{Nitish~Shirish Keskar}, \bibinfo{person}{Tabarak Khan}, \bibinfo{person}{Logan Kilpatrick}, \bibinfo{person}{Jong~Wook Kim}, \bibinfo{person}{Christina Kim}, \bibinfo{person}{Yongjik Kim}, \bibinfo{person}{Jan~Hendrik Kirchner}, \bibinfo{person}{Jamie Kiros}, \bibinfo{person}{Matt Knight}, \bibinfo{person}{Daniel Kokotajlo}, \bibinfo{person}{Łukasz Kondraciuk}, \bibinfo{person}{Andrew Kondrich},
  \bibinfo{person}{Aris Konstantinidis}, \bibinfo{person}{Kyle Kosic}, \bibinfo{person}{Gretchen Krueger}, \bibinfo{person}{Vishal Kuo}, \bibinfo{person}{Michael Lampe}, \bibinfo{person}{Ikai Lan}, \bibinfo{person}{Teddy Lee}, \bibinfo{person}{Jan Leike}, \bibinfo{person}{Jade Leung}, \bibinfo{person}{Daniel Levy}, \bibinfo{person}{Chak~Ming Li}, \bibinfo{person}{Rachel Lim}, \bibinfo{person}{Molly Lin}, \bibinfo{person}{Stephanie Lin}, \bibinfo{person}{Mateusz Litwin}, \bibinfo{person}{Theresa Lopez}, \bibinfo{person}{Ryan Lowe}, \bibinfo{person}{Patricia Lue}, \bibinfo{person}{Anna Makanju}, \bibinfo{person}{Kim Malfacini}, \bibinfo{person}{Sam Manning}, \bibinfo{person}{Todor Markov}, \bibinfo{person}{Yaniv Markovski}, \bibinfo{person}{Bianca Martin}, \bibinfo{person}{Katie Mayer}, \bibinfo{person}{Andrew Mayne}, \bibinfo{person}{Bob McGrew}, \bibinfo{person}{Scott~Mayer McKinney}, \bibinfo{person}{Christine McLeavey}, \bibinfo{person}{Paul McMillan}, \bibinfo{person}{Jake McNeil}, \bibinfo{person}{David
  Medina}, \bibinfo{person}{Aalok Mehta}, \bibinfo{person}{Jacob Menick}, \bibinfo{person}{Luke Metz}, \bibinfo{person}{Andrey Mishchenko}, \bibinfo{person}{Pamela Mishkin}, \bibinfo{person}{Vinnie Monaco}, \bibinfo{person}{Evan Morikawa}, \bibinfo{person}{Daniel Mossing}, \bibinfo{person}{Tong Mu}, \bibinfo{person}{Mira Murati}, \bibinfo{person}{Oleg Murk}, \bibinfo{person}{David Mély}, \bibinfo{person}{Ashvin Nair}, \bibinfo{person}{Reiichiro Nakano}, \bibinfo{person}{Rajeev Nayak}, \bibinfo{person}{Arvind Neelakantan}, \bibinfo{person}{Richard Ngo}, \bibinfo{person}{Hyeonwoo Noh}, \bibinfo{person}{Long Ouyang}, \bibinfo{person}{Cullen O'Keefe}, \bibinfo{person}{Jakub Pachocki}, \bibinfo{person}{Alex Paino}, \bibinfo{person}{Joe Palermo}, \bibinfo{person}{Ashley Pantuliano}, \bibinfo{person}{Giambattista Parascandolo}, \bibinfo{person}{Joel Parish}, \bibinfo{person}{Emy Parparita}, \bibinfo{person}{Alex Passos}, \bibinfo{person}{Mikhail Pavlov}, \bibinfo{person}{Andrew Peng}, \bibinfo{person}{Adam
  Perelman}, \bibinfo{person}{Filipe de Avila Belbute~Peres}, \bibinfo{person}{Michael Petrov}, \bibinfo{person}{Henrique~Ponde de Oliveira~Pinto}, \bibinfo{person}{Michael}, \bibinfo{person}{Pokorny}, \bibinfo{person}{Michelle Pokrass}, \bibinfo{person}{Vitchyr~H. Pong}, \bibinfo{person}{Tolly Powell}, \bibinfo{person}{Alethea Power}, \bibinfo{person}{Boris Power}, \bibinfo{person}{Elizabeth Proehl}, \bibinfo{person}{Raul Puri}, \bibinfo{person}{Alec Radford}, \bibinfo{person}{Jack Rae}, \bibinfo{person}{Aditya Ramesh}, \bibinfo{person}{Cameron Raymond}, \bibinfo{person}{Francis Real}, \bibinfo{person}{Kendra Rimbach}, \bibinfo{person}{Carl Ross}, \bibinfo{person}{Bob Rotsted}, \bibinfo{person}{Henri Roussez}, \bibinfo{person}{Nick Ryder}, \bibinfo{person}{Mario Saltarelli}, \bibinfo{person}{Ted Sanders}, \bibinfo{person}{Shibani Santurkar}, \bibinfo{person}{Girish Sastry}, \bibinfo{person}{Heather Schmidt}, \bibinfo{person}{David Schnurr}, \bibinfo{person}{John Schulman}, \bibinfo{person}{Daniel Selsam},
  \bibinfo{person}{Kyla Sheppard}, \bibinfo{person}{Toki Sherbakov}, \bibinfo{person}{Jessica Shieh}, \bibinfo{person}{Sarah Shoker}, \bibinfo{person}{Pranav Shyam}, \bibinfo{person}{Szymon Sidor}, \bibinfo{person}{Eric Sigler}, \bibinfo{person}{Maddie Simens}, \bibinfo{person}{Jordan Sitkin}, \bibinfo{person}{Katarina Slama}, \bibinfo{person}{Ian Sohl}, \bibinfo{person}{Benjamin Sokolowsky}, \bibinfo{person}{Yang Song}, \bibinfo{person}{Natalie Staudacher}, \bibinfo{person}{Felipe~Petroski Such}, \bibinfo{person}{Natalie Summers}, \bibinfo{person}{Ilya Sutskever}, \bibinfo{person}{Jie Tang}, \bibinfo{person}{Nikolas Tezak}, \bibinfo{person}{Madeleine~B. Thompson}, \bibinfo{person}{Phil Tillet}, \bibinfo{person}{Amin Tootoonchian}, \bibinfo{person}{Elizabeth Tseng}, \bibinfo{person}{Preston Tuggle}, \bibinfo{person}{Nick Turley}, \bibinfo{person}{Jerry Tworek}, \bibinfo{person}{Juan Felipe~Cerón Uribe}, \bibinfo{person}{Andrea Vallone}, \bibinfo{person}{Arun Vijayvergiya}, \bibinfo{person}{Chelsea Voss},
  \bibinfo{person}{Carroll Wainwright}, \bibinfo{person}{Justin~Jay Wang}, \bibinfo{person}{Alvin Wang}, \bibinfo{person}{Ben Wang}, \bibinfo{person}{Jonathan Ward}, \bibinfo{person}{Jason Wei}, \bibinfo{person}{CJ Weinmann}, \bibinfo{person}{Akila Welihinda}, \bibinfo{person}{Peter Welinder}, \bibinfo{person}{Jiayi Weng}, \bibinfo{person}{Lilian Weng}, \bibinfo{person}{Matt Wiethoff}, \bibinfo{person}{Dave Willner}, \bibinfo{person}{Clemens Winter}, \bibinfo{person}{Samuel Wolrich}, \bibinfo{person}{Hannah Wong}, \bibinfo{person}{Lauren Workman}, \bibinfo{person}{Sherwin Wu}, \bibinfo{person}{Jeff Wu}, \bibinfo{person}{Michael Wu}, \bibinfo{person}{Kai Xiao}, \bibinfo{person}{Tao Xu}, \bibinfo{person}{Sarah Yoo}, \bibinfo{person}{Kevin Yu}, \bibinfo{person}{Qiming Yuan}, \bibinfo{person}{Wojciech Zaremba}, \bibinfo{person}{Rowan Zellers}, \bibinfo{person}{Chong Zhang}, \bibinfo{person}{Marvin Zhang}, \bibinfo{person}{Shengjia Zhao}, \bibinfo{person}{Tianhao Zheng}, \bibinfo{person}{Juntang Zhuang},
  \bibinfo{person}{William Zhuk}, {and} \bibinfo{person}{Barret Zoph}.} \bibinfo{year}{2023}\natexlab{}.
\newblock \bibinfo{title}{GPT-4 Technical Report}.
\newblock
\newblock
\showeprint[arxiv]{2303.08774}~[cs.CL]


\bibitem[\protect\citeauthoryear{Owens}{Owens}{2023}]%
        {owens2023}
\bibfield{author}{\bibinfo{person}{Jim Owens}.} \bibinfo{year}{2023}\natexlab{}.
\newblock \bibinfo{booktitle}{\emph{Video Production Handbook} (\bibinfo{edition}{7th} ed.)}.
\newblock \bibinfo{publisher}{Routledge}.
\newblock
\urldef\tempurl%
\url{https://doi.org/10.4324/9781003251323}
\showURL{%
\tempurl}


\bibitem[\protect\citeauthoryear{Pandya and Holia}{Pandya and Holia}{2023}]%
        {pandya2023automatingcustomerserviceusing}
\bibfield{author}{\bibinfo{person}{Keivalya Pandya} {and} \bibinfo{person}{Mehfuza Holia}.} \bibinfo{year}{2023}\natexlab{}.
\newblock \bibinfo{title}{Automating Customer Service using LangChain: Building custom open-source GPT Chatbot for organizations}.
\newblock
\newblock
\showeprint[arxiv]{2310.05421}~[cs.CL]
\urldef\tempurl%
\url{https://arxiv.org/abs/2310.05421}
\showURL{%
\tempurl}


\bibitem[\protect\citeauthoryear{Park, O'Brien, Cai, Morris, Liang, and Bernstein}{Park et~al\mbox{.}}{2023}]%
        {park2023generative}
\bibfield{author}{\bibinfo{person}{Joon~Sung Park}, \bibinfo{person}{Joseph~C. O'Brien}, \bibinfo{person}{Carrie~J. Cai}, \bibinfo{person}{Meredith~Ringel Morris}, \bibinfo{person}{Percy Liang}, {and} \bibinfo{person}{Michael~S. Bernstein}.} \bibinfo{year}{2023}\natexlab{}.
\newblock \bibinfo{title}{Generative Agents: Interactive Simulacra of Human Behavior}.
\newblock
\newblock
\showeprint[arxiv]{2304.03442}~[cs.HC]


\bibitem[\protect\citeauthoryear{Park, Popowski, Cai, Morris, Liang, and Bernstein}{Park et~al\mbox{.}}{2022}]%
        {park2022social}
\bibfield{author}{\bibinfo{person}{Joon~Sung Park}, \bibinfo{person}{Lindsay Popowski}, \bibinfo{person}{Carrie Cai}, \bibinfo{person}{Meredith~Ringel Morris}, \bibinfo{person}{Percy Liang}, {and} \bibinfo{person}{Michael~S. Bernstein}.} \bibinfo{year}{2022}\natexlab{}.
\newblock \showarticletitle{Social Simulacra: Creating Populated Prototypes for Social Computing Systems}. In \bibinfo{booktitle}{\emph{Proceedings of the 35th Annual ACM Symposium on User Interface Software and Technology}} (Bend, OR, USA) \emph{(\bibinfo{series}{UIST '22})}. \bibinfo{publisher}{Association for Computing Machinery}, \bibinfo{address}{New York, NY, USA}, Article \bibinfo{articleno}{74}, \bibinfo{numpages}{18}~pages.
\newblock
\showISBNx{9781450393201}
\urldef\tempurl%
\url{https://doi.org/10.1145/3526113.3545616}
\showDOI{\tempurl}


\bibitem[\protect\citeauthoryear{Pavel, Goldman, Hartmann, and Agrawala}{Pavel et~al\mbox{.}}{2016}]%
        {videocrit2016pavel}
\bibfield{author}{\bibinfo{person}{Amy Pavel}, \bibinfo{person}{Dan~B. Goldman}, \bibinfo{person}{Bj\"{o}rn Hartmann}, {and} \bibinfo{person}{Maneesh Agrawala}.} \bibinfo{year}{2016}\natexlab{}.
\newblock \showarticletitle{VidCrit: Video-based Asynchronous Video Review}. In \bibinfo{booktitle}{\emph{Proceedings of the 29th Annual Symposium on User Interface Software and Technology}} (Tokyo, Japan) \emph{(\bibinfo{series}{UIST '16})}. \bibinfo{publisher}{Association for Computing Machinery}, \bibinfo{address}{New York, NY, USA}, \bibinfo{pages}{517–528}.
\newblock
\showISBNx{9781450341899}
\urldef\tempurl%
\url{https://doi.org/10.1145/2984511.2984552}
\showDOI{\tempurl}


\bibitem[\protect\citeauthoryear{Piro, Bianchi, Alessandrelli, Chizzola, Casiraghi, Sancassani, and Gatti}{Piro et~al\mbox{.}}{2024}]%
        {Piro2024MyLearningTalk}
\bibfield{author}{\bibinfo{person}{Ludovica Piro}, \bibinfo{person}{Tommaso Bianchi}, \bibinfo{person}{Luca Alessandrelli}, \bibinfo{person}{Andrea Chizzola}, \bibinfo{person}{Daniela Casiraghi}, \bibinfo{person}{Susanna Sancassani}, {and} \bibinfo{person}{Nicola Gatti}.} \bibinfo{year}{2024}\natexlab{}.
\newblock \showarticletitle{MyLearningTalk: An LLM-Based Intelligent Tutoring System}. In \bibinfo{booktitle}{\emph{Web Engineering}}, \bibfield{editor}{\bibinfo{person}{Kostas Stefanidis}, \bibinfo{person}{Kari Syst{\"a}}, \bibinfo{person}{Maristella Matera}, \bibinfo{person}{Sebastian Heil}, \bibinfo{person}{Haridimos Kondylakis}, {and} \bibinfo{person}{Elisa Quintarelli}} (Eds.). \bibinfo{publisher}{Springer Nature Switzerland}, \bibinfo{address}{Cham}, \bibinfo{pages}{428--431}.
\newblock
\showISBNx{978-3-031-62362-2}


\bibitem[\protect\citeauthoryear{Radford, Kim, Xu, Brockman, McLeavey, and Sutskever}{Radford et~al\mbox{.}}{2022}]%
        {whisper}
\bibfield{author}{\bibinfo{person}{Alec Radford}, \bibinfo{person}{Jong~Wook Kim}, \bibinfo{person}{Tao Xu}, \bibinfo{person}{Greg Brockman}, \bibinfo{person}{Christine McLeavey}, {and} \bibinfo{person}{Ilya Sutskever}.} \bibinfo{year}{2022}\natexlab{}.
\newblock \bibinfo{title}{Robust Speech Recognition via Large-Scale Weak Supervision}.
\newblock
\newblock
\showeprint[arxiv]{2212.04356}~[eess.AS]


\bibitem[\protect\citeauthoryear{Ramos and Balakrishnan}{Ramos and Balakrishnan}{2003}]%
        {Ramos2003annotateVideo}
\bibfield{author}{\bibinfo{person}{Gonzalo Ramos} {and} \bibinfo{person}{Ravin Balakrishnan}.} \bibinfo{year}{2003}\natexlab{}.
\newblock \showarticletitle{Fluid interaction techniques for the control and annotation of digital video}. In \bibinfo{booktitle}{\emph{Proceedings of the 16th Annual ACM Symposium on User Interface Software and Technology}} (Vancouver, Canada) \emph{(\bibinfo{series}{UIST '03})}. \bibinfo{publisher}{Association for Computing Machinery}, \bibinfo{address}{New York, NY, USA}, \bibinfo{pages}{105–114}.
\newblock
\showISBNx{1581136366}
\urldef\tempurl%
\url{https://doi.org/10.1145/964696.964708}
\showDOI{\tempurl}


\bibitem[\protect\citeauthoryear{Shao, Li, Dai, and Qiu}{Shao et~al\mbox{.}}{2023}]%
        {shao2023characterllmtrainableagentroleplaying}
\bibfield{author}{\bibinfo{person}{Yunfan Shao}, \bibinfo{person}{Linyang Li}, \bibinfo{person}{Junqi Dai}, {and} \bibinfo{person}{Xipeng Qiu}.} \bibinfo{year}{2023}\natexlab{}.
\newblock \bibinfo{title}{Character-LLM: A Trainable Agent for Role-Playing}.
\newblock
\newblock
\showeprint[arxiv]{2310.10158}~[cs.CL]
\urldef\tempurl%
\url{https://arxiv.org/abs/2310.10158}
\showURL{%
\tempurl}


\bibitem[\protect\citeauthoryear{Shen, Li, Su, Li, Chen, Jiang, and Xue}{Shen et~al\mbox{.}}{2017}]%
        {shen2017weakly}
\bibfield{author}{\bibinfo{person}{Zhiqiang Shen}, \bibinfo{person}{Jianguo Li}, \bibinfo{person}{Zhou Su}, \bibinfo{person}{Minjun Li}, \bibinfo{person}{Yurong Chen}, \bibinfo{person}{Yu-Gang Jiang}, {and} \bibinfo{person}{Xiangyang Xue}.} \bibinfo{year}{2017}\natexlab{}.
\newblock \showarticletitle{Weakly supervised dense video captioning}. In \bibinfo{booktitle}{\emph{Proceedings of the IEEE Conference on Computer Vision and Pattern Recognition}}. \bibinfo{pages}{1916--1924}.
\newblock


\bibitem[\protect\citeauthoryear{Siersdorfer, Chelaru, Nejdl, and San~Pedro}{Siersdorfer et~al\mbox{.}}{2010}]%
        {Siersdorfer2010predictcomment}
\bibfield{author}{\bibinfo{person}{Stefan Siersdorfer}, \bibinfo{person}{Sergiu Chelaru}, \bibinfo{person}{Wolfgang Nejdl}, {and} \bibinfo{person}{Jose San~Pedro}.} \bibinfo{year}{2010}\natexlab{}.
\newblock \showarticletitle{How useful are your comments? analyzing and predicting youtube comments and comment ratings}. In \bibinfo{booktitle}{\emph{Proceedings of the 19th International Conference on World Wide Web}} (Raleigh, North Carolina, USA) \emph{(\bibinfo{series}{WWW '10})}. \bibinfo{publisher}{Association for Computing Machinery}, \bibinfo{address}{New York, NY, USA}, \bibinfo{pages}{891–900}.
\newblock
\showISBNx{9781605587998}
\urldef\tempurl%
\url{https://doi.org/10.1145/1772690.1772781}
\showDOI{\tempurl}


\bibitem[\protect\citeauthoryear{Smilevski, Lalkovski, and Madjarov}{Smilevski et~al\mbox{.}}{2018}]%
        {Smilevski_2018}
\bibfield{author}{\bibinfo{person}{Marko Smilevski}, \bibinfo{person}{Ilija Lalkovski}, {and} \bibinfo{person}{Gjorgji Madjarov}.} \bibinfo{year}{2018}\natexlab{}.
\newblock \bibinfo{booktitle}{\emph{Stories for Images-in-Sequence by Using Visual and Narrative Components}}.
\newblock \bibinfo{publisher}{Springer International Publishing}, \bibinfo{pages}{148–159}.
\newblock
\showISBNx{9783030008253}
\showISSN{1865-0937}
\urldef\tempurl%
\url{https://doi.org/10.1007/978-3-030-00825-3_13}
\showDOI{\tempurl}


\bibitem[\protect\citeauthoryear{Song, Guo, Gao, Li, Hanjalic, and Shen}{Song et~al\mbox{.}}{2018}]%
        {song2018deterministic}
\bibfield{author}{\bibinfo{person}{Jingkuan Song}, \bibinfo{person}{Yuyu Guo}, \bibinfo{person}{Lianli Gao}, \bibinfo{person}{Xuelong Li}, \bibinfo{person}{Alan Hanjalic}, {and} \bibinfo{person}{Heng~Tao Shen}.} \bibinfo{year}{2018}\natexlab{}.
\newblock \showarticletitle{From deterministic to generative: Multimodal stochastic RNNs for video captioning}.
\newblock \bibinfo{journal}{\emph{IEEE transactions on neural networks and learning systems}} \bibinfo{volume}{30}, \bibinfo{number}{10} (\bibinfo{year}{2018}), \bibinfo{pages}{3047--3058}.
\newblock


\bibitem[\protect\citeauthoryear{{SSA}}{{SSA}}{2023}]%
        {2023USAName}
\bibfield{author}{\bibinfo{person}{{SSA}}.} \bibinfo{year}{2023}\natexlab{}.
\newblock \bibinfo{title}{USA Baby Name Dataset}.
\newblock
\newblock
\urldef\tempurl%
\url{https://www.ssa.gov/OACT/babynames/limits.html}
\showURL{%
\tempurl}


\bibitem[\protect\citeauthoryear{Stevenson and Phakiti}{Stevenson and Phakiti}{2014}]%
        {STEVENSON201451}
\bibfield{author}{\bibinfo{person}{Marie Stevenson} {and} \bibinfo{person}{Aek Phakiti}.} \bibinfo{year}{2014}\natexlab{}.
\newblock \showarticletitle{The effects of computer-generated feedback on the quality of writing}.
\newblock \bibinfo{journal}{\emph{Assessing Writing}}  \bibinfo{volume}{19} (\bibinfo{year}{2014}), \bibinfo{pages}{51--65}.
\newblock
\showISSN{1075-2935}
\urldef\tempurl%
\url{https://doi.org/10.1016/j.asw.2013.11.007}
\showDOI{\tempurl}
\newblock
\shownote{Feedback in Writing: Issues and Challenges.}


\bibitem[\protect\citeauthoryear{Tan, Jiang, and Ngo}{Tan et~al\mbox{.}}{2011}]%
        {Tan2011AudioVisual}
\bibfield{author}{\bibinfo{person}{Chun~Chet Tan}, \bibinfo{person}{Yu-Gang Jiang}, {and} \bibinfo{person}{Chong-Wah Ngo}.} \bibinfo{year}{2011}\natexlab{}.
\newblock \showarticletitle{Towards textually describing complex video contents with audio-visual concept classifiers}. In \bibinfo{booktitle}{\emph{Proceedings of the 19th ACM International Conference on Multimedia}} (Scottsdale, Arizona, USA) \emph{(\bibinfo{series}{MM '11})}. \bibinfo{publisher}{Association for Computing Machinery}, \bibinfo{address}{New York, NY, USA}, \bibinfo{pages}{655–658}.
\newblock
\showISBNx{9781450306164}
\urldef\tempurl%
\url{https://doi.org/10.1145/2072298.2072411}
\showDOI{\tempurl}


\bibitem[\protect\citeauthoryear{Tang, Wang, LIU, Rao, Li, and Li}{Tang et~al\mbox{.}}{2021}]%
        {Tang2021CLIP4Caption}
\bibfield{author}{\bibinfo{person}{Mingkang Tang}, \bibinfo{person}{Zhanyu Wang}, \bibinfo{person}{Zhenhua LIU}, \bibinfo{person}{Fengyun Rao}, \bibinfo{person}{Dian Li}, {and} \bibinfo{person}{Xiu Li}.} \bibinfo{year}{2021}\natexlab{}.
\newblock \showarticletitle{CLIP4Caption: CLIP for Video Caption}. In \bibinfo{booktitle}{\emph{Proceedings of the 29th ACM International Conference on Multimedia}} (Virtual Event, China) \emph{(\bibinfo{series}{MM '21})}. \bibinfo{publisher}{Association for Computing Machinery}, \bibinfo{address}{New York, NY, USA}, \bibinfo{pages}{4858–4862}.
\newblock
\showISBNx{9781450386517}
\urldef\tempurl%
\url{https://doi.org/10.1145/3474085.3479207}
\showDOI{\tempurl}


\bibitem[\protect\citeauthoryear{Thelwall, Sud, and Vis}{Thelwall et~al\mbox{.}}{2012}]%
        {thelwall2012commenting}
\bibfield{author}{\bibinfo{person}{Mike Thelwall}, \bibinfo{person}{Pardeep Sud}, {and} \bibinfo{person}{Farida Vis}.} \bibinfo{year}{2012}\natexlab{}.
\newblock \showarticletitle{Commenting on YouTube videos: From Guatemalan rock to el big bang}.
\newblock \bibinfo{journal}{\emph{Journal of the American society for information science and technology}} \bibinfo{volume}{63}, \bibinfo{number}{3} (\bibinfo{year}{2012}), \bibinfo{pages}{616--629}.
\newblock


\bibitem[\protect\citeauthoryear{{Thematic Analysis Inc.}}{{Thematic Analysis Inc.}}{2024}]%
        {thematic2024}
\bibfield{author}{\bibinfo{person}{{Thematic Analysis Inc.}}} \bibinfo{year}{2024}\natexlab{}.
\newblock \bibinfo{title}{Thematic Comment Analysis}.
\newblock
\newblock
\urldef\tempurl%
\url{https://getthematic.com/product/comment-analyzer/}
\showURL{%
\tempurl}
\newblock
\shownote{Accessed: 2024-10-01.}


\bibitem[\protect\citeauthoryear{Tseng, Huang, Hsiao, Chen, Huang, Meng, and Chen}{Tseng et~al\mbox{.}}{2024}]%
        {tseng2024talespersonallmssurvey}
\bibfield{author}{\bibinfo{person}{Yu-Min Tseng}, \bibinfo{person}{Yu-Chao Huang}, \bibinfo{person}{Teng-Yun Hsiao}, \bibinfo{person}{Wei-Lin Chen}, \bibinfo{person}{Chao-Wei Huang}, \bibinfo{person}{Yu Meng}, {and} \bibinfo{person}{Yun-Nung Chen}.} \bibinfo{year}{2024}\natexlab{}.
\newblock \bibinfo{title}{Two Tales of Persona in LLMs: A Survey of Role-Playing and Personalization}.
\newblock
\newblock
\showeprint[arxiv]{2406.01171}~[cs.CL]
\urldef\tempurl%
\url{https://arxiv.org/abs/2406.01171}
\showURL{%
\tempurl}


\bibitem[\protect\citeauthoryear{Van~Daele, Iyer, Zhang, Derry, Huh, and Pavel}{Van~Daele et~al\mbox{.}}{2024}]%
        {10.1145/3613904.3642839}
\bibfield{author}{\bibinfo{person}{Tess Van~Daele}, \bibinfo{person}{Akhil Iyer}, \bibinfo{person}{Yuning Zhang}, \bibinfo{person}{Jalyn~C Derry}, \bibinfo{person}{Mina Huh}, {and} \bibinfo{person}{Amy Pavel}.} \bibinfo{year}{2024}\natexlab{}.
\newblock \showarticletitle{Making Short-Form Videos Accessible with Hierarchical Video Summaries}. In \bibinfo{booktitle}{\emph{Proceedings of the 2024 CHI Conference on Human Factors in Computing Systems}} (Honolulu, HI, USA) \emph{(\bibinfo{series}{CHI '24})}. \bibinfo{publisher}{Association for Computing Machinery}, \bibinfo{address}{New York, NY, USA}, Article \bibinfo{articleno}{895}, \bibinfo{numpages}{17}~pages.
\newblock
\showISBNx{9798400703300}
\urldef\tempurl%
\url{https://doi.org/10.1145/3613904.3642839}
\showDOI{\tempurl}


\bibitem[\protect\citeauthoryear{{VEED}}{{VEED}}{2024}]%
        {veed2024}
\bibfield{author}{\bibinfo{person}{{VEED}}.} \bibinfo{year}{2024}\natexlab{}.
\newblock \bibinfo{title}{VEED}.
\newblock
\newblock
\urldef\tempurl%
\url{https://www.veed.io/}
\showURL{%
\tempurl}
\newblock
\shownote{Accessed: 2024-10-01.}


\bibitem[\protect\citeauthoryear{Wang, Li, Lv, Xia, Xu, and Sodhi}{Wang et~al\mbox{.}}{2024a}]%
        {wang2024lave}
\bibfield{author}{\bibinfo{person}{Bryan Wang}, \bibinfo{person}{Yuliang Li}, \bibinfo{person}{Zhaoyang Lv}, \bibinfo{person}{Haijun Xia}, \bibinfo{person}{Yan Xu}, {and} \bibinfo{person}{Raj Sodhi}.} \bibinfo{year}{2024}\natexlab{a}.
\newblock \bibinfo{title}{LAVE: LLM-Powered Agent Assistance and Language Augmentation for Video Editing}.
\newblock
\newblock
\showeprint[arxiv]{2402.10294}~[cs.HC]


\bibitem[\protect\citeauthoryear{Wang, Li, Lv, Xia, Xu, and Sodhi}{Wang et~al\mbox{.}}{2024b}]%
        {wang2024lavellmpoweredagentassistance}
\bibfield{author}{\bibinfo{person}{Bryan Wang}, \bibinfo{person}{Yuliang Li}, \bibinfo{person}{Zhaoyang Lv}, \bibinfo{person}{Haijun Xia}, \bibinfo{person}{Yan Xu}, {and} \bibinfo{person}{Raj Sodhi}.} \bibinfo{year}{2024}\natexlab{b}.
\newblock \bibinfo{title}{LAVE: LLM-Powered Agent Assistance and Language Augmentation for Video Editing}.
\newblock
\newblock
\showeprint[arxiv]{2402.10294}~[cs.HC]
\urldef\tempurl%
\url{https://arxiv.org/abs/2402.10294}
\showURL{%
\tempurl}


\bibitem[\protect\citeauthoryear{Wang, Yao, Kwok, and Ni}{Wang et~al\mbox{.}}{2020}]%
        {wang2020generalizing}
\bibfield{author}{\bibinfo{person}{Yaqing Wang}, \bibinfo{person}{Quanming Yao}, \bibinfo{person}{James~T Kwok}, {and} \bibinfo{person}{Lionel~M Ni}.} \bibinfo{year}{2020}\natexlab{}.
\newblock \showarticletitle{Generalizing from a few examples: A survey on few-shot learning}.
\newblock \bibinfo{journal}{\emph{ACM computing surveys (csur)}} \bibinfo{volume}{53}, \bibinfo{number}{3} (\bibinfo{year}{2020}), \bibinfo{pages}{1--34}.
\newblock


\bibitem[\protect\citeauthoryear{Wei, Wang, Schuurmans, Bosma, Xia, Chi, Le, Zhou, et~al\mbox{.}}{Wei et~al\mbox{.}}{2022}]%
        {wei2022chain}
\bibfield{author}{\bibinfo{person}{Jason Wei}, \bibinfo{person}{Xuezhi Wang}, \bibinfo{person}{Dale Schuurmans}, \bibinfo{person}{Maarten Bosma}, \bibinfo{person}{Fei Xia}, \bibinfo{person}{Ed Chi}, \bibinfo{person}{Quoc~V Le}, \bibinfo{person}{Denny Zhou}, {et~al\mbox{.}}} \bibinfo{year}{2022}\natexlab{}.
\newblock \showarticletitle{Chain-of-thought prompting elicits reasoning in large language models}.
\newblock \bibinfo{journal}{\emph{Advances in neural information processing systems}}  \bibinfo{volume}{35} (\bibinfo{year}{2022}), \bibinfo{pages}{24824--24837}.
\newblock


\bibitem[\protect\citeauthoryear{Wu, Jones, and Pitie}{Wu et~al\mbox{.}}{2020}]%
        {wu2020responselivebotgeneratinglive}
\bibfield{author}{\bibinfo{person}{Hao Wu}, \bibinfo{person}{Gareth J.~F. Jones}, {and} \bibinfo{person}{Francois Pitie}.} \bibinfo{year}{2020}\natexlab{}.
\newblock \bibinfo{title}{Response to LiveBot: Generating Live Video Comments Based on Visual and Textual Contexts}.
\newblock
\newblock
\showeprint[arxiv]{2006.03022}~[cs.CL]
\urldef\tempurl%
\url{https://arxiv.org/abs/2006.03022}
\showURL{%
\tempurl}


\bibitem[\protect\citeauthoryear{Yasunaga, Leskovec, and Liang}{Yasunaga et~al\mbox{.}}{2022}]%
        {yasunaga2022linkbert}
\bibfield{author}{\bibinfo{person}{Michihiro Yasunaga}, \bibinfo{person}{Jure Leskovec}, {and} \bibinfo{person}{Percy Liang}.} \bibinfo{year}{2022}\natexlab{}.
\newblock \showarticletitle{Linkbert: Pretraining language models with document links}.
\newblock \bibinfo{journal}{\emph{arXiv preprint arXiv:2203.15827}} (\bibinfo{year}{2022}).
\newblock


\bibitem[\protect\citeauthoryear{Yoon, Chen, Guimbreti\`{e}re, and Sellen}{Yoon et~al\mbox{.}}{2014}]%
        {yoon2014richreview}
\bibfield{author}{\bibinfo{person}{Dongwook Yoon}, \bibinfo{person}{Nicholas Chen}, \bibinfo{person}{Fran\c{c}ois Guimbreti\`{e}re}, {and} \bibinfo{person}{Abigail Sellen}.} \bibinfo{year}{2014}\natexlab{}.
\newblock \showarticletitle{RichReview: blending ink, speech, and gesture to support collaborative document review}. In \bibinfo{booktitle}{\emph{Proceedings of the 27th Annual ACM Symposium on User Interface Software and Technology}} (Honolulu, Hawaii, USA) \emph{(\bibinfo{series}{UIST '14})}. \bibinfo{publisher}{Association for Computing Machinery}, \bibinfo{address}{New York, NY, USA}, \bibinfo{pages}{481–490}.
\newblock
\showISBNx{9781450330695}
\urldef\tempurl%
\url{https://doi.org/10.1145/2642918.2647390}
\showDOI{\tempurl}


\bibitem[\protect\citeauthoryear{{YouTube}}{{YouTube}}{2023}]%
        {2023madeonyoutube}
\bibfield{author}{\bibinfo{person}{{YouTube}}.} \bibinfo{year}{2023}\natexlab{}.
\newblock \bibinfo{title}{Made on YouTube}.
\newblock
\newblock
\urldef\tempurl%
\url{https://blog.youtube/news-and-events/made-on-youtube-2023/}
\showURL{%
\tempurl}


\bibitem[\protect\citeauthoryear{Zeng, Chen, Chuang, Liao, Niebles, and Sun}{Zeng et~al\mbox{.}}{2017}]%
        {Zeng2017VideoQA}
\bibfield{author}{\bibinfo{person}{Kuo-Hao Zeng}, \bibinfo{person}{Tseng-Hung Chen}, \bibinfo{person}{Ching-Yao Chuang}, \bibinfo{person}{Yuan-Hong Liao}, \bibinfo{person}{Juan~Carlos Niebles}, {and} \bibinfo{person}{Min Sun}.} \bibinfo{year}{2017}\natexlab{}.
\newblock \showarticletitle{Leveraging Video Descriptions to Learn Video Question Answering}.
\newblock \bibinfo{journal}{\emph{Proceedings of the AAAI Conference on Artificial Intelligence}} \bibinfo{volume}{31}, \bibinfo{number}{1} (\bibinfo{date}{Feb.} \bibinfo{year}{2017}).
\newblock
\urldef\tempurl%
\url{https://doi.org/10.1609/aaai.v31i1.11238}
\showDOI{\tempurl}


\bibitem[\protect\citeauthoryear{Zeng, Gao, Xue, and Tu}{Zeng et~al\mbox{.}}{2021}]%
        {Zeng2021PLVCG}
\bibfield{author}{\bibinfo{person}{Zehua Zeng}, \bibinfo{person}{Neng Gao}, \bibinfo{person}{Cong Xue}, {and} \bibinfo{person}{Chenyang Tu}.} \bibinfo{year}{2021}\natexlab{}.
\newblock \showarticletitle{PLVCG: A Pretraining Based Model for Live Video Comment Generation}. In \bibinfo{booktitle}{\emph{Advances in Knowledge Discovery and Data Mining}}, \bibfield{editor}{\bibinfo{person}{Kamal Karlapalem}, \bibinfo{person}{Hong Cheng}, \bibinfo{person}{Naren Ramakrishnan}, \bibinfo{person}{R.~K. Agrawal}, \bibinfo{person}{P.~Krishna Reddy}, \bibinfo{person}{Jaideep Srivastava}, {and} \bibinfo{person}{Tanmoy Chakraborty}} (Eds.). \bibinfo{publisher}{Springer International Publishing}, \bibinfo{address}{Cham}, \bibinfo{pages}{690--702}.
\newblock
\showISBNx{978-3-030-75765-6}


\bibitem[\protect\citeauthoryear{Zhang, Yang, Feng, Qin, Chen, Yu, Chen, Wang, Savarese, Ermon, Xiong, and Xu}{Zhang et~al\mbox{.}}{2024}]%
        {Zhang_2024_CVPR}
\bibfield{author}{\bibinfo{person}{Shu Zhang}, \bibinfo{person}{Xinyi Yang}, \bibinfo{person}{Yihao Feng}, \bibinfo{person}{Can Qin}, \bibinfo{person}{Chia-Chih Chen}, \bibinfo{person}{Ning Yu}, \bibinfo{person}{Zeyuan Chen}, \bibinfo{person}{Huan Wang}, \bibinfo{person}{Silvio Savarese}, \bibinfo{person}{Stefano Ermon}, \bibinfo{person}{Caiming Xiong}, {and} \bibinfo{person}{Ran Xu}.} \bibinfo{year}{2024}\natexlab{}.
\newblock \showarticletitle{HIVE: Harnessing Human Feedback for Instructional Visual Editing}. In \bibinfo{booktitle}{\emph{Proceedings of the IEEE/CVF Conference on Computer Vision and Pattern Recognition (CVPR)}}. \bibinfo{pages}{9026--9036}.
\newblock


\bibitem[\protect\citeauthoryear{Zhang*, Kishore*, Wu*, Weinberger, and Artzi}{Zhang* et~al\mbox{.}}{2020}]%
        {bert-score}
\bibfield{author}{\bibinfo{person}{Tianyi Zhang*}, \bibinfo{person}{Varsha Kishore*}, \bibinfo{person}{Felix Wu*}, \bibinfo{person}{Kilian~Q. Weinberger}, {and} \bibinfo{person}{Yoav Artzi}.} \bibinfo{year}{2020}\natexlab{}.
\newblock \showarticletitle{BERTScore: Evaluating Text Generation with BERT}. In \bibinfo{booktitle}{\emph{International Conference on Learning Representations}}.
\newblock
\urldef\tempurl%
\url{https://openreview.net/forum?id=SkeHuCVFDr}
\showURL{%
\tempurl}


\bibitem[\protect\citeauthoryear{Zhu, Lu, Zheng, Guo, Zhang, Wang, and Yu}{Zhu et~al\mbox{.}}{2018}]%
        {zhu2018texygen}
\bibfield{author}{\bibinfo{person}{Yaoming Zhu}, \bibinfo{person}{Sidi Lu}, \bibinfo{person}{Lei Zheng}, \bibinfo{person}{Jiaxian Guo}, \bibinfo{person}{Weinan Zhang}, \bibinfo{person}{Jun Wang}, {and} \bibinfo{person}{Yong Yu}.} \bibinfo{year}{2018}\natexlab{}.
\newblock \showarticletitle{Texygen: A benchmarking platform for text generation models}. In \bibinfo{booktitle}{\emph{The 41st international ACM SIGIR conference on research \& development in information retrieval}}. \bibinfo{pages}{1097--1100}.
\newblock


\end{thebibliography}
